\documentclass[acmsmall]{acmart}

\AtBeginDocument{%
  }

\setcopyright{acmlicensed}
\copyrightyear{2018}
\acmYear{2018}
\acmDOI{XXXXXXX.XXXXXXX}

\acmJournal{JACM}
\acmVolume{37}
\acmNumber{4}
\acmArticle{111}
\acmMonth{8}

\usepackage{multirow} 
\usepackage{multicol}
\usepackage{algorithm}
\usepackage{algpseudocode}
\usepackage{subfigure}
\usepackage{enumitem}

\newcommand{\nosection}[1]{\vspace{2pt}\noindent\textbf{#1.}}

\newcommand{\lyy}[1]{\textcolor{black}{#1}}
\newcommand{\sigmoid}{\text{sigmoid}}  
\newtheorem{definition}{Definition}
  
\begin{document}

\title{Post-Training Attribute Unlearning in Recommender Systems}

\author{Chaochao Chen}
\email{zjuccc@zju.edu.cn}
\orcid{0000-0003-1419-964X}
\affiliation{%
  \institution{Zhejiang University}
  \city{Hangzhou}
  \country{China}
}

\author{Yizhao Zhang}
\email{22221337@zju.edu.cn}
\affiliation{%
  \institution{Zhejiang University}
  \city{Hangzhou}
  \country{China}}

\author{Yuyuan Li}
\email{y2li@hdu.edu.cn}
\authornote{Corresponding author.}
\affiliation{%
  \institution{Hangzhou Dianzi University}
  \city{Hangzhou}
  \country{China}
}

\author{Jun Wang}
\email{junwang.lu@gmail.com}
\affiliation{%
  \institution{OPPO Research Institute}
  \city{Shenzhen}
  \country{China}
  }

\author{Lianyong Qi}
\email{lianyongqi@upc.edu.cn}
\affiliation{
 \institution{China University of Petroleum}
 \city{Qingdao}
 \country{China}}

 \author{Xiaolong Xu}
 \email{njuxlxu@gmail.com}
\affiliation{
 \institution{Nanjing University of Information Science and Technology}
 \city{Nanjing}
 \country{China}}
 
\author{Xiaolin Zheng}
\email{xlzheng@zju.edu.cn}
\affiliation{%
 \institution{Zhejiang University}
 \city{Hangzhou}
 \country{China}}

\author{Jianwei Yin}
\email{zjuyjw@cs.zju.edu.cn}
\affiliation{%
 \institution{Zhejiang University}
 \city{Hangzhou}
 \country{China}}

\renewcommand{\shortauthors}{Chaochao Chen, et al.}

\begin{abstract}
  With the growing privacy concerns in recommender systems, recommendation unlearning is getting increasing attention. Existing studies predominantly use training data, i.e., model inputs, as unlearning target.  However, attackers can extract private information from the model even if it has not been explicitly encountered during training. We name this unseen information as \textit{attribute} and treat it as unlearning target. To protect the sensitive attribute of users, Attribute Unlearning (AU) aims to make target attributes indistinguishable. In this paper, we focus on a strict but practical setting of AU, namely Post-Training Attribute Unlearning (PoT-AU), where unlearning can only be performed after the training of the recommendation model is completed. To address the PoT-AU problem in recommender systems, we propose a two-component loss function. The first component is distinguishability loss, where we design a distribution-based measurement to make attribute labels indistinguishable from attackers. We further extend this measurement to handle multi-class attribute cases with efficient computational overhead. The second component is regularization loss, where we explore a function-space measurement that effectively maintains recommendation performance compared to parameter-space regularization. We use stochastic gradient descent algorithm to optimize our proposed loss. Extensive experiments on four real-world datasets demonstrate the effectiveness of our proposed methods.
\end{abstract}



\received{20 February 2007}
\received[revised]{12 March 2009}
\received[accepted]{5 June 2009}

\maketitle

\section{Introduction}\label{sec:intro}
To alleviate the issue of information overload~\cite{isinkaye2015recommendation, zhao2024funnelrag}, recommender systems have been widely applied in practice with great success, having a substantial influence on people's lifestyles~\cite{schafer2007collaborative, han2023intra, chen2022differential, 10707460}.
The success lies in their ability to extract highly personalized information from user data.
However, people have grown more aware of privacy concerns in personalized recommendations, and demand their sensitive information be protected.
As one of the protective measures, \textit{Right to be Forgotten}~\cite{2014gdpr, 2018ccpa, 2019pipeda} requires recommendation platforms to enable users to withdraw their individual data and its impact, which impulses the study of machine/recommendation unlearning.

\begin{figure}[t]
    \centering
    \includegraphics[width=10cm]{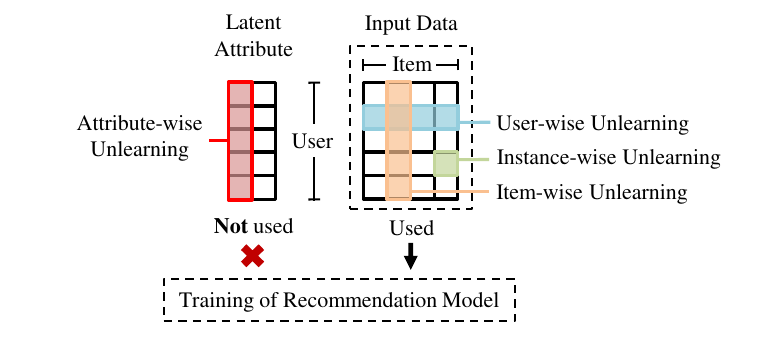}
    \caption{Illustrations of different unlearning targets.}
    \label{fig:setting}
\end{figure}

Existing studies on machine unlearning mainly use training data, i.e., model inputs, as the unlearning target~\cite{nguyen2022survey}.
We name this type of unlearning task as Input Unlearning (IU). 
As shown in Fig.~\ref{fig:setting}, in the recommendation scenarios, the input data can be a user-item interaction matrix.
With different unlearning targets, IU can be user-wise, item-wise, and instance-wise~\cite{chen2022recommendation}.
IU benefits multiple parties, e.g., data providers and model owners, because the target data can be i) the specified data that contains users' sensitive information, and ii) the dirty data that is polluted by accidental mistakes or intentional attack~\cite{li2016data}.

\begin{table} 
    \centering
    \caption{Difference between input unlearning and attribute unlearning in recommender systems.}
    \resizebox{0.68\linewidth}{!}{
    \begin{tabular}{c|c|c}
         \toprule
           &  \textbf{Input Unlearning} & \textbf{Attribute Unlearning}\\
         \midrule
         \multirow{2}{*}{Unlearning target} & Input data & Latent attribute\\
         & (used in training) & (\textbf{not} used in training)\\
         \midrule
         Applicability of & \multirow{2}{*}{Ground truth} & \multirow{2}{*}{Not applicable}\\
         retraining from scratch & \\
         \bottomrule
    \end{tabular}
    }
    \label{tab:type}
\end{table}


Extensive studies on IU cannot obscure the importance of Attribute Unlearning (AU), where attributes represent the inherent properties, e.g., gender, race, and age of users that have \textbf{not} been used for training (Table~\ref{tab:type}: difference in unlearning target) but implicitly learned by embedding models.
Due to the information extraction capabilities of recommender systems, AU is especially valuable in the context of recommendation.
Although recommendation models did not see the latent attribute, existing research has found that basic machine learning models can successfully infer users' attributes from the user embeddings learned by collaborative filtering models~\cite{ganhor2022unlearning}, which is also known as attribute inference attack~\cite{jia2018attriguard, beigi2020privacy, zhang2021graph, zhang2023comprehensive}.
Therefore, from the perspective of privacy preservation, AU is as important as IU in recommender systems.
However, existing IU methods cannot be applied in AU.
As illustrated in Table~\ref{tab:type}, retraining from scratch (ground truth for IU) is unable to unlearn the latent attribute, i.e., not applicable for AU, since it is not explicitly utilized during training at all.

Existing but limited research on AU has focused on In-Training AU (InT-AU)~\cite{guo2022efficient,ganhor2022unlearning}, where unlearning is performed during model training (as shown in the right part of Fig.~\ref{fig:overview}).
In this paper, we focus on a more strict AU setting, namely Post-Training Attribute Unlearning (PoT-AU), where we can only manipulate the model after training and have no knowledge about training data or other training information (as shown in the left part of Fig.~\ref{fig:overview}
).
Compared with InT-AU, this setting is more strict, because of \textit{data accessibility}, i.e., we may not get access to the training data or other information after training due to regulations.
PoT-AU is also more practical than InT-AU, because of \textit{deployment overhead}, i.e., non-interference with the original training process is more flexible and reduces deployment overhead.
%
%
As shown in Fig.~\ref{fig:overview}, there are two goals for PoT-AU in recommender systems. 
The primary goal (\textbf{Goal \#1}) is to make the target attribute indistinguishable to the inference attacking. 
The other goal (\textbf{Goal \#2}) is to maintain the recommendation performance, as both users and recommendation platforms want to avoid harming the original recommendation tasks.

To achieve the above two goals in the PoT-AU problem, Li~et~al.~\cite{li2023making} consider it as an optimization problem concerning user embeddings. They subsequently design a two-component loss function that consists of distinguishability loss and regularization loss. 
Although effective for the PoT-AU problem, this method only considers binary-class attributes, neglecting the more common multi-class attributes found in real-world scenarios. This oversight reduces the practical applicability of the PoT-AU method. In the context of multi-class attributes, this method has two major shortcomings.
Firstly, the distinguishability loss was designed to minimize the distance between two groups of user embeddings, which leads to significant computational complexity for multi-class attributes, especially when the number of label categories is large.
Secondly, we observed that the performance of recommendation decreases when attribute unlearning is performed, particularly in the multi-class scenario. 
This decline in performance can be attributed to the discrepancy between the proposed parameter-space regularization loss~\cite{li2023making} and the intended function-space regularization, as evidenced by our empirical study in Section~\ref{sec:compare}.
Analyzing the above two shortcomings, we identify two key challenges for PoT-AU, \textbf{CH1}: How can we reduce the computational complexity of multi-class attribute unlearning? \textbf{CH2}: How can we maintain the recommendation performance while achieving attribute unlearning?

\textbf{Our work.} To address these challenges for multi-class attributes, we further modify the design of both distinguishability loss and regularization loss.
For \textbf{CH1}, we establish an \textit{anchor} distribution and minimize the distance between other distributions with it. This approach reduces the computational complexity from \textit{$O(T^2)$} to \textit{$O(T)$}, where $T$ is the number of attribute categories, e.g., female and male when $T = 2$.
For \textbf{CH2}, we propose a data-free regularization loss $\ell_r$ in the function space, which directly regularizes the function of the model to preserve recommendation performance. \lyy{This approach enhances the effectiveness of regularization compared to traditional $\ell_2$ loss in parameter space.}

\textbf{Our contributions.} It is worth mentioning that this work is an extension of our previous work~\cite{li2023making}. Compared with~\cite{li2023making}, we extend the
study of binary-class attributes to the multi-class scenario, identifying the shortcomings of our previous work in this scenario, i.e., \textit{significant computational complexity} and \textit{limited preservation of recommendation performance}. 
%
To overcome these two shortcomings, we i) establish an anchor distribution to mitigate computational complexity, and ii) propose a data-free regularization loss in the function space to directly align recommendation performance. As will be shown in Fig.~\ref{fig:corr}, there is a negative correlation between our proposed regularization loss and the similarity between recommendation performance before and after unlearning. This correlation indicates that our regularization loss is more effective than the $\ell_2$ loss proposed in~\cite{li2023making}. Furthermore, we conduct additional experiments of AU in the multi-class scenario to demonstrate the effectiveness and efficiency of our proposed method.
\lyy{To the best of our knowledge, this is the first work to explore the multi-class scenario in attribute unlearning, thereby enhancing the overall completeness and real-world applicability of our previous research.}
We summarize the main contributions of this paper as follows:

%
%

\begin{itemize}[leftmargin=*]\setlength{\itemsep}{-\itemsep}
    \item Following our previous work~\cite{li2023making}, we study the PoT-AU problem. 
    We identify two essential goals of PoT-AU, and propose a two-component loss function, with each component devised to target one of the aforementioned goals.
    \item To address \textbf{CH1}, we extend the distributional perspective distinguishability loss from binary-class attributes to the multi-class scenario by introducing an anchor distribution. 
    \item To address \textbf{CH2}, we explore a data-free function-space measurement as the regularization loss to maintain the recommendation performance during unlearning.
    \item We conduct extensive experiments on four real-world datasets with in-depth analyses to evaluate the effectiveness of our proposed methods regarding both unlearning (\textbf{Goal \#1}) and recommendation (\textbf{Goal \#2}).
    %
\end{itemize}


\section{Related Work}

In this section, in addition to AU, we also briefly introduce machine unlearning and recommendation unlearning to offer a comprehensive literature review.

\subsection{Machine Unlearning}
Machine unlearning, an emerging paradigm in the field of privacy-preserving machine learning, aims to completely remove user's data from a trained model~\cite{nguyen2022survey}.
A straightforward unlearning method is to retrain the model from scratch on the dataset that eliminates the target data. 
However, it is computationally prohibitive for large-scale models in real-world scenarios.
Current studies on machine unlearning can be divided into two main categories based on the level of unlearning completeness. 

\begin{itemize}[leftmargin=*] \setlength{\itemsep}{-\itemsep}
\item \textbf{Exact Unlearning} aims to ensure that the data is completely unlearned from the model, akin to retraining from scratch. 
Cao and Yang~\cite{cao2015towards} first studied the machine unlearning problem and transformed training data points into a reduced number of summations to enhance unlearning efficiency. 
Bourtoule~et~al.~\cite{bourtoule2021machine} proposed a general unlearning method, i.e. SISA (Sharded, Isolated, Sliced and Aggregated), based on partition-aggregation framework.
SISA reduces the retraining overhead to subsets. 
Recently, Yan~et~al.~\cite{yan2022arcane} proposed a novel partition-aggregation unlearning framework, i.e., ARCANE, which partitions data by class.
To enable training for each subset, ARCANE transforms the original classification task into multiple one-class classification tasks. 
\item \textbf{Approximate Unlearning} aims to estimate the influence of unlearning target, and directly remove the influence through parameter manipulation, i.e., updating parameters with the purpose of unlearning~\cite{golatkar2020eternal,guo2020certified,sekhari2021remember,warnecke2021machine}. 
Approximate unlearning relaxes the definition of exact unlearning and only provides a statistical guarantee of unlearning completeness.
The influence of target data is usually estimated by influence function~\cite{koh2017understanding, koh2019accuracy}.
However, it is found to be fragile in deep learning~\cite{basu2021influence}.
\end{itemize}

%
%

%
%

\subsection{Recommendation Unlearning}
Following SISA's partition-aggregation framework, Chen~et~al.~\cite{chen2022recommendation} 
proposed an exact recommendation unlearning framework named RecEraser, which groups similar data together and uses an attention-based aggregator to enhance recommendation performance.
Similarly, LASER also groups similar data together~\cite{li2022making}.
%
Lately, Li~et~al.~\cite{li2023ultrare} proposed a novel grouping method based on optimal transport theory to obtain partition results more effectively and efficiently.
%
Approximate unlearning is also investigated in the context of recommendation~\cite{li2023selective, zhang2023recommendation}.
A benchmark has been proposed to comprehensively evaluate various recommendation unlearning methods~\cite{CURE4Rec2024}.

\subsection{Attribute Unlearning}
Existing studies of machine unlearning predominately focus on unlearning specific samples from the training data, ignoring the latent attributes that are irrelevant to the training process.
Guo~et~al.~\cite{guo2022efficient} firstly studied the AU problem and proposed to manipulate disentangled representatives to unlearn particular attributes of facial images, e.g., smiling, mustache, and big nose. Specifically, the manipulation is achieved by splitting the model into a feature extractor and a classifier, and then adding a network block between them. 
Furthermore, Moon~et~al.~\cite{moon2023feature} investigated AU in generative models, e.g., generative adversarial nets and variational autoencoders, by learning a transformation from the image containing the target attribute to the image without it.

As recommender systems potentially capture the sensitive information of users, e.g., gender, race, and age, AU is non-trivial in the recommendation scenario.
However, representative manipulation and learning a transformation with public datasets may not be universally applicable in the context of recommendation ~\cite{guo2022efficient}. For AU in recommendation, 
Ganhor~et~al.~\cite{ganhor2022unlearning} introduce adversarial training to achieve AU for recommendation model based on variational autoencoder. This work is under the setting of In-Training AU (InT-AU), which involves manipulating the training process. 
Different from InT-AU, our previous work~\cite{li2023making} and this work aims to achieve \textit{model-agnostic} AU under the \textit{post-training} setting (PoT-AU). This is more challenging because 
i) we can only manipulate the model parameters when training is completed, 
and, ii) as the training data or other training information, e.g., gradients, are usually protected or discarded after training, we cannot get access to them to enhance performance.
At the same time, PoT-AU is more practical, because it is more flexible for recommendation platforms to manipulate the model based on unlearning requests without interfering with the original process of training. 

\section{Preliminaries\label{sec:pre}}

In this section, we first revisit the paradigm of collaborative filtering models. Then, we specify the details of attribute inference attack. The notations used in this paper are listed in Table~\ref{tab:notation}. 


\begin{table*} 
  \caption{Description of Notations}
  \label{tab:notation}
  \begin{tabular}{cl}
    \toprule
    Notations & Description \\
    \midrule
    $\mathcal{U}$, $\mathcal{V}$ & The set of users and items \\
    $M$, $N$                     & The number of users and items \\
    $u$, $v$                     & The user and item \\
    $\mathcal{R}$                & The set of interactions \\
    $\mathcal{R^-}$              & The set of sampled negative interactions \\
    $r_{u, v}$                   & The interaction between $u$ and $v$ \\
    $s_{u, v}$                   & The predicted score of recommendation model between $u$ and $v$ \\
    $\hat{s}_{u, v}$            & The predicted score of unlearned recommendation model between $u$ and $v$ \\
    $\mathbf{e}_u, \mathbf{e}_v$ & The embedding of user $u$ and item $v$ \\
    $\boldsymbol{\theta}$            & The user embedding matrix \\
    $\boldsymbol{\beta}$             & The weight of each distribution for computing anchor distribution \\
    $d$                          & The dimension of user embedding \\
    $S_i$                        & The i-th category of attribute \\
    $T$                          & The sum of categories of attribute \\
    $\mathbb{P}_i$               & The distribution of user embedding with label $S_i$ \\  
    $\mathcal{G}$                & The reproducing kernel Hilbert space with Gaussian kernel function \\
    $Dist$                       & The measure of discrepancy between distributions \\
    $sim$                        & The cosine similarity \\
    $k$                          & The length of top-$k$ item lists for ranking alignment \\
    $K$                          & The length of the recommendation list for NDCG and HR metric\\
    $\ell_2$                     & L2 regularization term \\
    $\ell_u$                     & The distinguishability loss \\
    $\ell_r$                     & functional regularization term \\
    $\lambda$                    & The margin in $\ell_r$ \\
    $\mathbf{w}$                 & The weight of margin in $\ell_r$ \\
    \bottomrule
  \end{tabular}
\end{table*}

\subsection{Collaborative Filtering}\label{sec:rec}
Discovering user preferences on items based on historical behavior forms the foundation of collaborative filtering modeling~\cite{shi2014collaborative,mnih2007probabilistic,hu2008collaborative}. Let $\mathcal{U}$ = \{$u_1, \dots, u_M$\} and $\mathcal{V}$ = \{$v_1, \dots, v_N$\} denote the user and item set, respectively. The interaction set $\mathcal{R} = \{(u,v) | \text{$u$ interacted with $v$}\}$ indicates the implicit relationships between each user in $\mathcal{U}$ and his/her consumed items. The interaction set $\mathcal{R} = \{(u,v) | \text{$u$ interacted with $v$}\}$ indicates the implicit interaction.
In general, many existing collaborative filtering approaches are designed with encoder network $f(\cdot)$ to generate low-dimensional representations of users and items $f(u), f(v) \in \mathbb{R}^d$ ($d$ is the dimension of latent space). For example, matrix factorization models typically employ an embedding table as the encoder, while graph-based models incorporate neighborhood information into the encoder. Then, the predicted score is defined as the similarity between user and item representation (e.g., dot product). Regarding the learning objective, most studies adopt the Bayesian Personalized Ranking (BPR)~\cite{rendle2012bpr} loss or the Cross Entropy (CE) loss~\cite{he2017neural} to train the model:
\begin{equation}\label{bpr_loss}
\mathcal{L}_{BPR} = \frac{1}{\lvert \mathcal{R} \rvert} \sum_{(u,v) \in \mathcal{R}} -\log(\sigmoid(s_{u,v}-s_{u,v^-})),
\end{equation}  
\begin{align}\label{ce_loss}
\mathcal{L}_{CE} = & \frac{1}{\lvert \mathcal{R \cup R^-} \rvert} \sum_{(u,v) \in \mathcal{R \cup R^-}} r_{u,v} \log(s_{u,v})  + (1 - r_{u,v})\log(1 - s_{u,v}),
\end{align}  
where $v^-$ is a randomly sampled negative item that the user has not interacted with, $\mathcal{R}^-$ is the set of negative samples, $s$ denotes the predicted score. $r_{u,v}$ denotes the interaction between $u$ and $v$, which is set as 1 if $(u,v) \in \mathcal{R}$ and 0 otherwise.

\subsection{Attacking Setting}

The process of attacking in PoT-AU problem is also known as the attribute inference attack, which poses a significant threat to both users and models. This attack can also be an evaluation metric to assess the effectiveness of attribute unlearning, an approach we adopt in our experiments. Specifically, the attack process consists of three stages, i.e., exposure, training, and inference.
%
%
In the \textit{exposure} stage, we assume that attackers follow the setting of grey-box attacks.
In other words, not all model parameters but only users' embeddings and their corresponding attribute information are exposed to attackers.
In the \textit{training} stage, we assume that attackers train the attacking model on a shadow dataset, which can be generated by sampling from the original users or users from the same distribution~\cite{salem2018ml}.
Although shadow-dataset training will inevitably reduce attacking performance, this assumption is reasonable, since the full-dataset setting is too strong and impractical.
Note that in our experiment, to ensure the reliability and validity of the evaluation, we construct an attacker using 80\% of users as the shadow dataset to enhance the performance of the attacker, and we perform five-fold cross-validation.
Regarding the attack as a classification task, the attacker uses user embeddings as input data and attribute information as labels.
Different from~\cite{li2023making}, we extend the binary setting to multi-class scenarios in this paper.
In the \textit{inference} stage, attackers use their trained attacking models to make predictions.

Note that our paper adopts a different attacking setting compared to previous studies on defense against attribute inference attack~\cite{beigi2020privacy, zhang2021graph, zhang2023comprehensive}. 
Specifically, our focus in attacking is primarily on the privacy of trained models rather than the implicit information presented in the original interaction data, aligning with the goal of attribute unlearning. This is because access to training data is limited within the context of PoT-AU. 
Additionally, instead of using the top-$k$ recommended item list (model output), we select the embedding layer of collaborative filtering model as the input for the attacking model. 


\section{Post-Training Attribute Unlearning}

In this section, we provide a detailed explanation of our motivation and delve into the process of the PoT-AU problem in recommender systems.
Subsequently, we consider the PoT-AU problem as an optimization problem 
and propose a novel two-component loss function to address it.

\subsection{Motivation}  
\begin{figure*}[t]
    \centering
    \includegraphics[width=\textwidth]{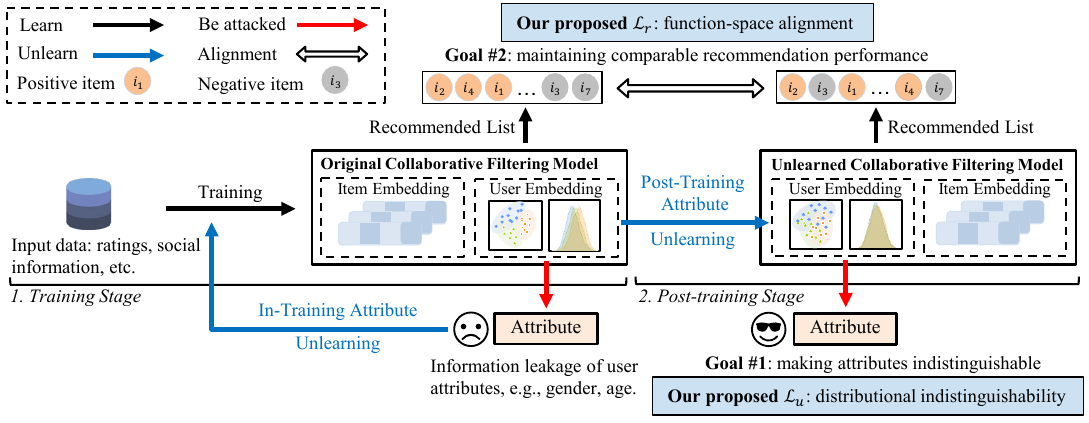}
    \caption{An overview of Post-Training Attribute Unlearning (PoT-AU) vs In-Training Attribute Unlearning (InT-AU) in recommender systems. $\mathcal{L}_u$ denotes the distinguishability loss designed for \textbf{Goal \#1}, $\mathcal{L}_r$ denotes the regularization loss designed for \textbf{Goal \#2}. The orange dots represent positive items which are in the top-$l$ positions of recommended list, while the gray dots represent the opposite. 
    We omit other parameters in the collaborative filtering model besides embeddings for conciseness.}
    \label{fig:overview}
\end{figure*}

As shown in Fig.~\ref{fig:overview}, we divide the entire process of PoT-AU into two stages, i.e., the training stage and the post-training stage.
In the training stage, the recommender system trains an original collaborative filtering model using input data.
To align with the post-training setting, we leave this stage untamed and assume that no additional information in this stage is available, except for the recommendation model and the attributes of users.
In the post-training stage, we generate new user embedding by unlearning the original one. 
The updated embeddings, i.e., user embeddings after unlearning, are supposed to achieve two goals simultaneously.
\begin{itemize}[leftmargin=*] \setlength{\itemsep}{-\itemsep}
    \item \textbf{Goal \#1} (unlearning) is to make target attributes distinguishable so as to protect attribute information from attackers.
    \item \textbf{Goal \#2} (recommendation) is to maintain the original recommendation performance, ensuring that the initial requirements of users are not compromised.
\end{itemize}

Compared with the In-Training (InT) setting, the Post-Training (PoT) setting is more challenging.
Firstly, PoT-AU allows no interference with the training process. Adding network block~\cite{guo2022efficient}, and adversarial training~\cite{ganhor2022unlearning} are not applicable under this setting.
Secondly, even though PoT-AU cuts down the connection with the training process, directly manipulating user embeddings by adding artificially designed noise, e.g., differential privacy~\cite{abadi2016deep}, is inappropriate. because i) it will inevitably degrade recommendation performance, and ii) its unlearning ability is not promising, as the functional mechanism of attacking models, including complex machine learning models, is not well understood. 
Thirdly, PoT-AU prohibits access to the input data and other training information that could be either unavailable or under protection and cannot be used for fine-tuning user embeddings, e.g., adding noise to the embeddings and then fine-tuning to boost recommendation performance.

In this paper, we further extend our previous study of binary-class attributes to the more multi-class scenario, which holds broader applicability in practice. The motivation for this extension stems from addressing two key challenges outlined in Section~\ref{sec:intro}, i.e., \textbf{CH1} (high computational complexity) and \textbf{CH2} (compromise in recommendation performance), which arise from directly applying our previous work to the multi-class scenario.

\subsection{Two-Component Loss Function \label{method:two}}
In the context of the PoT setting, one feasible solution is to conceptualize the desired final user embeddings while temporarily disregarding the intermediate manipulation and transformation processes. As a result, we formulate the PoT-AU as an optimization problem on user embeddings. In other words, our aim is to devise a suitable loss function and leverage optimization techniques to accomplish the task. Our previous work has demonstrated the effectiveness of this approach~\cite{li2023making}.

Specifically, we propose a two-component loss function that is specifically devised to address the two goals in the PoT-AU problem, i.e., \textbf{Goal \#1}: unlearning and \textbf{Goal \#2}: recommendation. 
Each component of the loss function is tailored to achieve one of these goals.
The trade-off coefficient $\alpha$ is introduced to get a balance between attribute unlearning and recommendation: 
\begin{equation}\label{equ:loss}
    L(\boldsymbol{\theta}) = \ell_u + \alpha \ell_r,
\end{equation}
where $\boldsymbol{\theta} \in \mathbb{R}^{M\times d}$ denotes user embeddings to be updated, $\ell_u$ and $\ell_r$ represent the loss for \textbf{Goal \#1} and \textbf{Goal \#2} respectively.

\subsection{Distinguishability Loss}
The core difficulty of designing a proper two-component loss function lies in defining distinguishability loss $\ell_u$, which is related to the primary goal of PoT-AU, i.e., \textbf{Goal \#1}: making the target attribute indistinguishable.
In our previous work, we define the distinguishability from a perspective of distribution, namely Distribution-to-Distribution loss (D2D)~\cite{li2023making}. 
Without loss of generality, we assume the target attribute has binary labels: $S_1$ and $S_2$, and extend it to multi-class scenarios in Section~\ref{sec:mul}.

\subsubsection{Binary-Class Scenario} \label{sec:mmd_binary}
We consider the user embeddings with the same attribute label as a distribution, e.g., $\mathbb{P}_1$ denotes the embedding distribution of users with label $S_1$.
For practical consideration, it is worth noting that the embeddings of all users are trained together without any attribution information. 
As a result, the shapes of the embedding distribution tend to be similar across different attribute labels.
The difference in distributions mainly comes from their distance.
Therefore, we use distributional distance $dist(\mathbb{P}_1, \mathbb{P}_2)$ to measure distinguishability.
We name this type of distinguishability measurement as D2D loss and define it as follows:
\begin{definition}[Distribution-to-Distribution Distinguishability~\cite{li2023making}]\label{def:d2d}
    Given two distributions of embedding from users with different attribute labels $P_{\theta_1}$ and $P_{\theta_2}$, we define distribution-to-distribution distinguishability as the distance between two distributions:
    \begin{equation} \label{equ:l_ud}
        \ell_{u, D} = Dist(\mathbb{P}_1, \mathbb{P}_2).
    \end{equation}
\end{definition}
%
Here, we apply MMD with radial kernels~\cite{tolstikhin2016minimax} to measure the distance of two distributions, which satisfies several properties that are required as a distance measurement, including non-negativity and exchange invariance, i.e., $Dist(\mathbb{P}_1, \mathbb{P}_2) = Dist(\mathbb{P}_2, \mathbb{P}_1)$. 
Specifically, by mapping the original distributions to a reproducing kernel Hilbert space $\mathcal{G}$ with function $\phi(\cdot)$, the MMD between $\mathbb{P}_1$ and $\mathbb{P}_2$ is defined as:
\begin{equation}\label{equ:att_ori}
    \text{MMD}^2(\mathbb{P}_1, \mathbb{P}_2) = \sup_{\|\phi\|_\mathcal{\mathcal{G}}\leq 1}\|\mathbb{E}_{\boldsymbol{\theta}_1\sim\mathbb{P}_1}[\phi(\boldsymbol{\theta}_1)] - \mathbb{E}_{\boldsymbol{\theta}_2\sim\mathbb{P}_2}[\phi(\boldsymbol{\theta}_2)]\|_\mathcal{G}^2,
\end{equation}
where $\mathbb{E}_{\boldsymbol{\theta}_1\sim\mathbb{P}_1}[\cdot]$ denotes the expectation with regard to distribution $\mathbb{P}_{1}$ in $\mathcal{G}$, i.e., kernel mean embedding, $\|\phi\|_\mathcal{G}\leq1$ defines a set of functions in the unit ball of $\mathcal{G}$. For simplicity, we let $\mu$ to denote kernel mean embedding of the distribution $\mathbb{P}$, then we have $\mu(\mathbb{P}) = \int \phi(\theta) d \mathbb{P(\theta)} $. Given a collection of samples $\boldsymbol{\theta} = \{\theta_1, . . . , \theta_n\}$, a natural empirical estimator~\cite{sriperumbudur2010hilbert, chatalic2022nystrom} of kernel mean embedding is given by:
\begin{equation} \label{equ:KME_esti}
    \mu({\mathbb{P}}) = \frac{1}{n} \sum_{i=1}^{n} \phi(\theta_i).
\end{equation}
Thus, given $n$ samples from $\boldsymbol{\theta}_1\sim\mathbb{P}_1$ and $m$ samples from $\boldsymbol{\theta}_2\sim\mathbb{P}_2$, MMD can be empirically estimated~\cite{gretton2012kernel} as:
\begin{align}
    & \hat{\text{MMD}^2}(\mathbb{P}_1, \mathbb{P}_2) = \frac{1}{n(n-1)}\sum_{i=1}^n\sum_{j\neq i}^n\kappa(\boldsymbol{\theta}_1^i, \boldsymbol{\theta}_1^j) + \frac{1}{m(m-1)}\sum_{i=1}^m\sum_{j\neq i}^m\kappa(\boldsymbol{\theta}_2^i, \boldsymbol{\theta}_2^j) - \frac{2}{nm}\sum_{i=1}^n\sum_{j=1}^m\kappa(\boldsymbol{\theta}_1^i, \boldsymbol{\theta}_2^j),
\end{align}
where $\kappa(\cdot, \cdot)$ is the kernel function, i.e., Gaussian kernel function~\cite{scholkopf1997comparing}. Based on MMD, we have the distinguishability loss $\ell_u$:

\begin{equation}\label{equ:loss_lu_binary}
    \ell_u = \text{MMD}^2(\mathbb{P}_1, \mathbb{P}_2).
\end{equation}

\subsubsection{Multi-Class Scenario}\label{sec:mul}
Given that the computational complexity of MMD in binary-class scenarios is assumed to be $O(1)$, minimizing $\ell_{u, D}$ for each pair of ($\mathbb{P}_1$, $\mathbb{P}_2$) can become computationally prohibitive in multi-class scenarios with a large number of label categories, i.e., $T$.
In such cases, the computational complexity increases to $O(T^2)$. 
Moreover, note that directly minimizing $\ell_{u, D}$ of each distribution pair may lead to instability during unlearning.

To extend our proposed $\ell_{u, D}$ loss to multi-class attribute unlearning, we introduce an \textit{anchor distribution} to reduce complexity.  
Specifically, given $T$ distributions, the anchor distribution is defined as a distribution $\mathbb{P}^{*}$, which minimizes the average sum of weighted distances between itself and and the aforementioned $T$ distributions. 
This objective is equivalent to identifying an interpolation between several probability measures, which is also known as barycenter estimation~\cite{agueh2011barycenters}. Formally, we have:
\begin{equation}
    \mathbb{P}^{*} = {\arg \min}_{\mathbb{P}}{\sum_{i=1}^T \beta_i \cdot Dist(\mathbb{P}, \mathbb{P}_i)},
\end{equation}
where $\mathbb{P}$ denotes an interpolation distribution of user embedding, and $\beta_i$ denotes the weight of distribution $\mathbb{P}_{i}$. 
The weight $\beta_i$ is typically determined empirically based on the size of the distribution, i.e., $|\mathbb{P}_i|/M$~\cite{montesuma2021wasserstein, silvia2020general}.

Previous studies~\cite{agueh2011barycenters, cuturi2014fast} introduce Wasserstein distance to compute barycenter. 
However, within the context of PoT-AU, the computational complexity of estimating Wasserstein barycenter grows exponentially when the dimension of user embedding $d$ increases. 
Therefore, following our choice in binary-class scenarios (Section~\ref{sec:mmd_binary}), we use the MMD distance with Gaussian kernel to estimate the barycenter for simplicity \lyy{and consistency}. Specifically, we have:
%
%
\begin{equation}
    \mathbb{P}^{*} = {\arg \min}_{\mathbb{P}}{\sum_{i=1}^T \beta_i \|[\mu(\mathbb{P}) - \mu(\mathbb{P}_{i})]\|_\mathcal{G}^2},
\end{equation}
which is equivalent to finding an optimal kernel mean embedding $\mu^*$ in $\mathcal{H}$ that minimizes
\begin{equation}\label{equ:mu}
    \mu^{*} = {\arg \min}_{\mu \in \mathcal{G}} {\sum_{i=1}^T \beta_i \|[\mu - \mu(\mathbb{P}_{i})]\|_\mathcal{G}^2}.
\end{equation}
As Equation~(\ref{equ:mu}) is a strongly convex quadratic function of $\mu$, the minimum is given by the first-order condition:
\begin{equation}
    \mu^{*} = \sum_{i=1}^T \beta_i \mu(\mathbb{P}_i).
\end{equation}
As the integral in kernel mean embedding is estimated by Equation~(\ref{equ:KME_esti}), we can set $\beta_i = |\mathbb{P}_i|/M$ to obtain:
\begin{align} \label{equ:optimal_mixture}
    & \mu^{*} = \sum_{i=1}^T \beta_i \mu(\mathbb{P}_i) =  \mu(\sum_{i=1}^T \beta_i \mathbb{P}_i), \nonumber\\
    & \mathbb{P}^{*} = \sum_{i=1}^T \beta_i \mathbb{P}_i.
\end{align}
Thus, we can obtain the anchor distribution by weighted interpolation, i.e., Equation~(\ref{equ:optimal_mixture}). 
For the implementation, we perform sampling from the distribution of all user embeddings to estimate the anchor distribution $\mathbb{P}^*$ without extra computational cost.

With the help of anchor distribution, we can reduce the computational complexity of $\ell_u$ from $O(T^2)$ to $O(T)$ by only calculating the MMD distance between the anchor distribution and the $T$ distributions. Formally, we have:
\begin{equation}\label{equ:loss_mul_u}
    \ell_u = \frac{1}{T} \sum_{i=1}^T \text{MMD}^2(\mathbb{P}_i, \mathbb{P}^{*}).
\end{equation}
With our proposed D2D distinguishability loss $\ell_u$, we can not only preserve the shape of user embedding distributions, but also efficiently achieve attribute unlearning in multi-class scenarios.


\subsection{Regularization Loss}
To achieve \textbf{Goal \#2} under the PoT setting, we introduce a data-free regularization loss, namely $\ell_r$, in Equation~(\ref{equ:loss}). 
This is necessary as we lack access to training data, and therefore can only rely on regularization loss to maintain recommendation performance while conducting unlearning.

\subsubsection{Regularization in Parameter Space} 
\label{exp:regular}
In previous work~\cite{li2023making}, we employ the widely acknowledged $\ell_2$ norm~\cite{bottcher2008frobenius} as the regularization loss, which regularizes user embedding in the parameter space.
This approach is based on the intuition that closer model parameters will lead to similar model performance, thus preserving the recommendation performance.
Formally, we have:
\begin{equation}
    \ell_2 = \|\boldsymbol{\theta} - \boldsymbol{\theta}^*\|_F^2 = \sum_{i=1}^{M}\sum_{j=1}^d(\theta_{i, j} - \theta^*_{i, j})^2,
\end{equation}
where $\boldsymbol{\theta}^*$ denotes the original user embeddings before unlearning. 

\subsubsection{Regularization in Function Space}
However, this intuition may be inaccurate during model training and fine-tuning.
Benjamin~et~al.~\cite{benjamin2018measuring} found that a change in the parameter space might serve as a \textit{poor indicator} for the change in the function space, i.e., model performance. 
 
Similar to the scenario of PoT-AU, continual learning requires optimizing the model without utilizing training data while maintaining performance on the original task. 
Motivated by previous studies in continual learning~\cite{li2017learning, rannen2017encoder, kang2022class}, we consider a more fundamental regularization method, i.e., functional regularization, to achieve \textbf{Goal\#2} without accessing training data. The function of recommendation models is to provide users with a list of recommended items by mining their preferences, thus we fetch the recommended list before unlearning as the target of regularization.
Given that items positioned at the top of rank lists hold greater significance compared to those lower down~\cite{tang2018ranking}, we only regularize the top-$k$ recommended items for each user. Specifically, we formulate the regularization of rank list as a learning-to-rank task, and introduce a data-free rank regularization loss, denoted as $\ell_r$. Instead of regularizing user embeddings in parameter space, we focus on minimizing the discrepancy in the order of top-$k$ items in the recommended list before and after unlearning. This approach directly regularizes user embeddings in function space, aligning perfectly with \textbf{Goal\#2}.

Here we use the pair-wise loss to regularize the original top-$k$ item list~\cite{burges2005learning, reddi2021rankdistil, zhu2023membership}.
Formally, we have:
\begin{align}\label{equ:reg_lr}
    & \ell_{pr} = \sum_{i=1}^{M}\bigg[\sum_{j=1}^{k-1}\max(0, \hat{s}_{u_i,v^{i}_{j+1}}-\hat{s}_{u_i, v^{i}_{j}}+\lambda_1) + \sum_{j=1}^{k}\max(0,\hat{s}_{u_i, neg^{i}_{j}}-\hat{s}_{u_i, v^{i}_{j}} + \lambda_2)\bigg],
\end{align} 
where $v^{i}_{j}$ denotes the $j$-th item in the top-$k$ list of user $u_{i}$ before unlearning, and $\hat{s}$ denotes the predicted score between user and item after unlearning. 
We also sample $k$ items that are not in the original top-$k$ list of user $u_{i}$ as negative samples, where $neg^{i}_j$ denotes the $j$-th negative item of user $u_{i}$ (without consideration of order). $\lambda_1$ and $\lambda_2$ are two margin values, which are regarded as hyper-parameters. This loss function is composed of two pairwise terms based on hinge loss~\cite{gentile1998linear}. 
The first term aims to maximize the probability of ranking positive items in the same order as the top-$k$ list before unlearning, while the second term aims to improve the score of items in the top-$k$ list. 
However, directly regularizing the unlearning optimization with $\ell_{pr}$ may have a negative impact on the recommendation performance. $\ell_{pr}$ only considers the relative order of the items in the first k positions, but ignores the absolute difference between them. Since $\lambda_1$ and $\lambda_2$ are fixed, $\ell_{pr}$ may amplify the rating difference between similar items and reduce the rating difference between dissimilar items. To solve this problem, we propose an adaptive weight for $\lambda$. Specifically, we assume that the weight of margin for an item pair $(v_i,v_{i+1})$ should be negatively correlated to the similarity between $v_i$ and $v_{i+1}$:
\begin{align}\label{equ:weight}
    & w_{v_i,v_{i+1}} \propto \frac{1}{sim(\mathbf{e}_{v_i},\mathbf{e}_{v_{i+1}})},
\end{align} 
where $sim(\cdot)$ denotes the cosine similarity between item embeddings. Following~\cite{reddi2021rankdistil}, we use a parametrized geometric distribution for weighting the margin:


\begin{align}\label{equ:weight_exp}
    & w_{v_i,v_{i+1}} \propto 1 - \sigmoid[\frac{sim(\mathbf{e}_{v_i}, \mathbf{e}_{v_{i+1}})}{\tau}],
\end{align} 
where $\tau$ denotes the hyper-parameter that controls the sharpness of the distribution. Finally, we have:
\begin{align}\label{equ:reg_lr_adaptive}
    \ell_{r} & = \sum_{i=1}^{M}\bigg[\sum_{j=1}^{k-1}\max(0, \hat{s}_{u_i,v^{i}_{j+1}}-\hat{s}_{u_i, v^{i}_{j}} + w_{v^{i}_j,v^{i}_{j+1}} \cdot \lambda) + \sum_{j=1}^{k}\max(0,\hat{s}_{u_i, neg^{i}_{j}}-\hat{s}_{u_i, v^{i}_{j}} + w_{v^{i}_{j}, neg^{i}_j}  \cdot \lambda)\bigg]\nonumber\\
    & = \sum_{i=1}^{M}\bigg[\sum_{j=1}^{k-1}\max_{\text{pos}} + \sum_{j=1}^{k}\max_{\text{neg}} \bigg].
\end{align}

By utilizing $\ell_r$, we can more directly and effectively maintain the model's performance while conducting unlearning.

\begin{figure}[t]
\centering
\subfigure[\textbf{$\ell_2$ loss (perturbation)}]{\includegraphics[height=3.4cm,width=4.25cm]{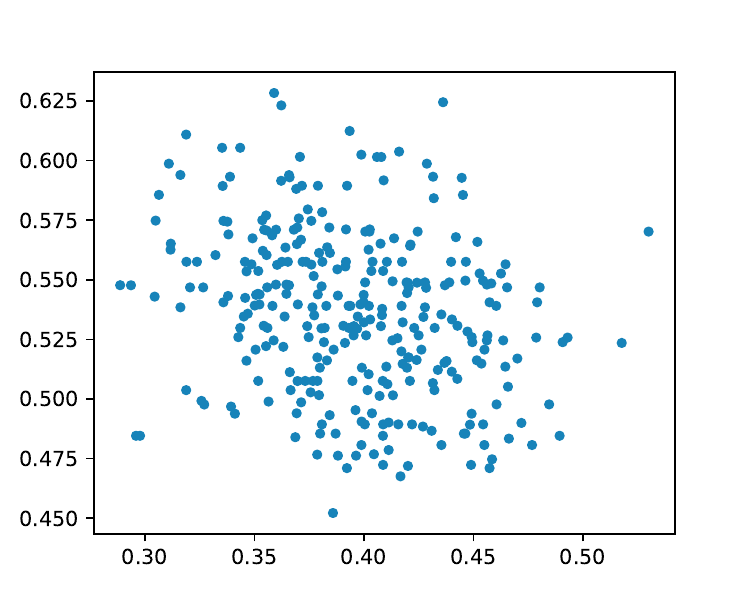} \label{fig:l2_cor}}
\subfigure[\textbf{$\ell_r$ loss (perturbation)}]{\includegraphics[height=3.4cm,width=4.25cm]{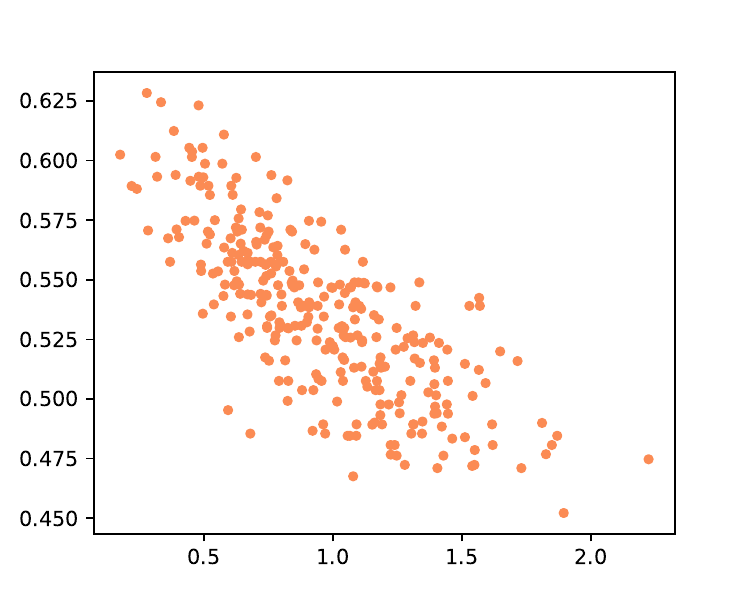} \label{fig:kd_cor}}
\caption{Correlation between two types of regularization losses and RBO (similarity in recommendation performance), where the x-axis and y-axis represent values of losses and RBO, respectively.
Note that $\ell_2$ is a parameter-space regularization, and $\ell_r$ is a function-space regularization. 
(a) Adding perturbation and calculating $\ell_2$;
(b) Adding perturbation and calculating $\ell_r$.
The Pearson correlation coefficients for (a) and (b) are 
-0.255 and -0.766 respectively.}
\label{fig:corr_random}
\end{figure}

\subsubsection{Comparison of Parameter and Function Spaces}\label{sec:compare}
We conduct a simulated empirical study to investigate the discrepancy between parameter and function spaces in the context of PoT-AU.
Specifically, we directly add Gaussian perturbations into the original user embeddings to simulate random changes in parameters. This process is repeated 300 times to observe the discrepancy in regularization losses and recommendation performance, i.e., model function. 
We use Rank Biased Overlap (RBO)~\cite{webber2010similarity} to measure the similarity of top@10 recommended item lists, which reflects discrepancy in the function space.
Note that the perturbation budget is set as 0.5 ($\|\Delta_u\| \leq 0.5$, where $\Delta_u$ denotes the perturbation.)

Based on the visual results (Fig.~\ref{fig:corr_random}), it is evident that there is a substantial correlation between our newly proposed function-space regularization loss $\ell_r$ and RBO.
In contrast, the parameter-space regularization loss $\ell_2$ exhibits a relatively lower correlation with RBO. 
Specifically, the Pearson correlation coefficient for $\ell_r$ is -0.766, whereas for $\ell_2$, it is merely -0.255.
This observation provides evidence of the limited effectiveness of the parameter-space loss $\ell_2$ in accurately measuring the changes in the function space. 
However, our newly proposed function-space regularization loss $\ell_r$ shows a stronger capability in this regard, thereby contributing to the preservation of recommendation performance. To comprehensively evaluate the proposed $\ell_r$ loss, we also analyze the difference between regularization in the parameter space and function space during the attribute unlearning process in Section~\ref{exp:rq5}.

\subsection{Putting Together} 
Incorporating the proposed distinguishability loss $\ell_u$ (Equation~(\ref{equ:loss_mul_u})) and regularization loss $\ell_r$ (Equation~(\ref{equ:reg_lr_adaptive})), we formulate the two-component loss function (Equation~(\ref{equ:loss})) for the PoT-AU problem. This newly proposed loss function offers i) extra computational efficiency for multi-class attribute scenarios, and ii) superior preservation of recommendation performance.
Specifically, the loss function is computed by
\begin{equation}
    L_1(\boldsymbol{\theta}) = \underbrace{\frac{1}{T} \sum_{i=1}^T \text{MMD}^2(\mathbb{P}_i, \mathbb{P}^{*})}_{\ell_u} + \alpha \underbrace{\sum_{i=1}^{M}\bigg[\sum_{j=1}^{k-1}\max_{\text{pos}} + \sum_{j=1}^{k}\max_{\text{neg}} \bigg]}_{\ell_r}.
\end{equation}
Note that the loss function in our previous work is computed by
\begin{equation}
    L_2(\boldsymbol{\theta}) = \underbrace{\text{MMD}^2(\mathbb{P}_1, \mathbb{P}_2)}_{\ell_r} + \alpha \underbrace{\|\boldsymbol{\theta} - \boldsymbol{\theta}^*\|_F^2}_{\ell_r}.
\end{equation}
%
We apply the stochastic gradient descent algorithm~\cite{bottou2012stochastic} to optimize our proposed loss.
We investigate the effect of $\alpha$ and other hyper-parameters in Section~\ref{exp:rq4}.

\section{Experiments}

To comprehensively evaluate our proposed methods, we conduct experiments on four benchmark datasets and observe the performance in terms of unlearning and recommendation.
We also investigate the efficiency and robustness of our proposed loss functions.
We further conduct a detailed analysis of the unlearning process and compared D2D-PR with D2D-FR to showcase the superior effectiveness of D2D-FR in preserving recommendation performance. Specifically, We aim to answer the following research questions (RQs):

\begin{itemize}[leftmargin=*] \setlength{\itemsep}{-\itemsep}
    \item \textbf{RQ1}: Can our method effectively unlearning attributes under the setting of PoT-AU?
    \item \textbf{RQ2}: can our method maintain the recommendation performance after unlearning?
    \item \textbf{RQ3}: How about the efficiency of our proposed method?

    \item \textbf{RQ4}: What is the impact of key hyper-parameters in terms of unlearning and recommendation performance of our proposed method?
    \item \textbf{RQ5}: What is the contribution of our proposed D2D-FR compared with D2D-PR?
    \item \textbf{RQ6}: Can our method maintain unlearning performance when the attribute inference attacker utilizes different kinds of attacking models?
\end{itemize}

\subsection{Experimental Settings}

\subsubsection{Datasets}
Experiments are conducted on four publicly accessible datasets that contain both input data, i.e., user-item interactions, and user attributes, i.e., gender, age, and country.
\begin{itemize}[leftmargin=*] \setlength{\itemsep}{-\itemsep}
    \item \textbf{MovieLens 100K (ML-100K)}\footnote{https://grouplens.org/datasets/movielens/}: MovieLens is one of the most widely used datasets in the recommendation~\cite{harper2015movielens, he2016ups}.
    They collected users' ratings towards movies as well as other attributes, e.g., gender, age, and occupation.
    ML-100K is the version containing 100 thousand ratings. 
    
    \item \textbf{MovieLens 1M (ML-1M)}: A version of MovieLens dataset that has 1 million ratings.
    
    \item \textbf{LFM-2B}\footnote{http://www.cp.jku.at/datasets/LFM-2b}: This dataset collected more than 2 billion listening events, which is used for music retrieval and recommendation tasks~\cite{melchiorre2021investigating}. LFM-2B also contains user attributes including gender and country. Here we use a subset of the whole dataset which includes more than 3 million ratings.

    \item \textbf{KuaiSAR-small}\footnote{https://kuaisar.github.io/}: KuaiSAR is a unified search and recommendation dataset containing the genuine user behavior logs collected from the short-video mobile app, Kuaishou\footnote{https://www.kuaishou.com/}. Here we use a tiny version of KuaiSAR, i.e., KuaiSAR-small. It also includes two attributes of users, namely Feat1 and Feat2.


\end{itemize} 

For these datasets, we first filter out the users without valid attribute information, then we only keep the users that rated at least 5 items and the items with at least 5 user interactions following~\cite{xue2017deep, he2017neural}. The characteristics of datasets are summarized in Table~\ref{tab:dataset}.

To evaluate the recommendation performance, we use the leave-one-out method which is widely used in literature~\cite{he2017neural}. That is, we reserve the last two items for each user (ranked by the timestamp of interaction), one as the validation item and the other as the test item.

Regarding attribute data, we utilize three attributes, i.e., age, gender and country, from MovieLens and LFM-2B. Following~\cite{beigi2020privacy, zhang2021graph,ganhor2022unlearning}, we categorize the age attribute into three groups, i.e., over-45, under-35, and between 35 and 45, while the provided gender attribute is limited to females and males. As for KuaiSAR, we utilize the encrypted one-hot anonymous categories of users as the target attribute.

\begin{table}
\caption{Summary of datasets.}
\resizebox{0.68\linewidth}{!}{
\label{tab:dataset}
\begin{tabular}{lcccccc}  
\toprule
Dataset & Attribute & Category \# & User \#   & Item \#   & Rating \# & Sparsity \\
\midrule
\multirow{2}{*}{ML-100K} & Gender & 2 & \multirow{2}{*}{943} & \multirow{2}{*}{1,349}  & \multirow{2}{*}{99,287} & \multirow{2}{*}{92.195\%} \\
& Age & 3 & & & & \\
\midrule
\multirow{2}{*}{ML-1M}   & Gender & 2 & \multirow{2}{*}{6,040} & \multirow{2}{*}{3,416}     & \multirow{2}{*}{999,611} & \multirow{2}{*}{95.155\%} \\
& Age & 3 & & & &\\
\midrule
\multirow{2}{*}{LFM-2B}    & Gender & 2 & \multirow{2}{*}{19,972} & \multirow{2}{*}{99,639}   & \multirow{2}{*}{2,829,503}  & \multirow{2}{*}{99.858\%} \\
& Country & 8 & & & & \\
\midrule
\multirow{2}{*}{KuaiSAR}    & Feat1  & 7 & \multirow{2}{*}{21,852} & \multirow{2}{*}{140,367}   & \multirow{2}{*}{2,166,893}   & \multirow{2}{*}{99.929\%} \\
& Feat2 & 2 & & & \\

\bottomrule
\end{tabular}
}
\end{table}

\subsubsection{Evaluation Metrics}

\paragraph{\textbf{Attribute Unlearning Effectiveness}} 
As mentioned in Section~\ref{sec:rec}, we focus on collaborative filtering models and use \textit{user embeddings} as the attacking and unlearning target. Here we build a strong adversary classifier, i.e., attacker:
\begin{itemize}[leftmargin=*] \setlength{\itemsep}{-\itemsep}
    \item \textbf{MLP}~\cite{gardner1998artificial}: Multilayer Perceptron (MLP) is a simplified two-layer neural network, which is a widely used classifier. Here the dimension of hidden layer is set as 100 and a softmax layer is used as the output layer.
\end{itemize}
According to the previous study~\cite{li2023making}, MLP stands out as the attacker with best performance. We also investigate other types of attackers and different structures of MLP, with the results reported in Section~\ref{exp:rq6}, aligning consistently with the findings in \cite{li2023making}.
To quantify the effectiveness of model unlearning, we utilize two commonly used classification metrics: Micro-F1 Score (F1) and Balanced-Accuracy (BAcc) to evaluate the performance of attribute inference attack following~\cite{ganhor2022unlearning,grandini2020metrics}. Note that lower F1 scores and BAccs indicate better unlearning effectiveness. Following ~\cite{zhang2021graph,beigi2020privacy}, we use 80\% of the users to train the attacker, and the remainder for testing. The results of attribute inference attack are averaged over five runs using five-fold cross-validation. To ensure a fair comparison, we tune the hyper-parameters and optimize until the loss function converges, thus obtaining the optimal unlearning effectiveness.

\paragraph{\textbf{Recommendation Effectiveness}}
 To evaluate the performance of recommendation, we use the leave-one-out approach~\cite{he2016fast} to generate test samples. We leverage Hit Ratio at rank $K$ (HR@$K$) and Normalized Discounted Cumulative Gain at rank $K$ (NDCG@$K$) as measures of recommendation performance. HR@$K$ measures whether the test item is present in the top-$K$ list, while NDCG@$K$ are position-aware ranking metrics that assign higher scores to the hits at upper ranks~\cite{he2015trirank,xue2017deep}.
In our experiment, the entire negative item sets rather than the sampled subsets are used to compute HR@$K$ and NDCG@$K$, this is because the sampled metrics have been observed to be unstable and inconsistent when compared to their exact version~\cite{krichene2020sampled}. Note that we compare the recommendation performance of several methods under the condition of achieving the optimal unlearning effectiveness respectively.

\subsubsection{Recommendation Models}
We test our proposed methods on two different recommendation models:
\begin{itemize}[leftmargin=*] \setlength{\itemsep}{-\itemsep}
    \item \textbf{NMF}~\cite{he2017neural}: Neural Matrix Factorization (NMF) is one of the representative models based on matrix factorization.
    \item \textbf{LightGCN}~\cite{he2020lightgcn}: Light Graph Convolution Network (LightGCN) is the state-of-the-art collaborative filtering model which improves recommendation performance by simplifying the graph convolution network. 
\end{itemize}



\subsubsection{Unlearning Methods} 
Although the setting of InT-AU differs from that of PoT-AU, comparing our proposed methods with InT-AU approaches would contribute to a comprehensive understanding of the AU problem. Therefore, we compare our proposed methods with the original user embedding and two InT-AU methods.

\begin{itemize}[leftmargin=*] \setlength{\itemsep}{-\itemsep}
    \item \textbf{Original}: This is the original model before unlearning.
    \item \textbf{Retrain}~\cite{zafar2019fairness} (InT-AU): This method incorporates the aforementioned D2D loss into the original recommendation loss and retrains the model from scratch.
    \item \textbf{Adv-InT}~\cite{ganhor2022unlearning} (InT-AU): This method uses adversarial training to achieve InT-AU for the MultVAE~\cite{shenbin2020recvae}. We also apply the idea of adversarial training to our tested recommendation models, i.e., NMF and LightGCN, and name it Adv-InT.
    \item \textbf{D2D-PR}~\cite{li2023making} (PoT-AU): This is our previous work using a two-component loss function with D2D loss as distinguishability loss and $\ell_2$ as regularization loss.
    \item \textbf{D2D-FR} (PoT-AU): This is a two-component loss function with our newly proposed $\ell_u$ as distinguishability loss and $\ell_r$ as regularization loss, i.e., Equation~(\ref{equ:loss}).
\end{itemize}

\subsubsection{Parameter Settings and Hardware Information}

\begin{itemize}[leftmargin=*] \setlength{\itemsep}{-\itemsep}
    \item \textbf{Hardware Information}: All models and algorithms are implemented with Python 3.8 and PyTorch 1.9. We run all experiments
    on an Ubuntu 20.04 LTS System server with 256GB RAM and NVIDIA GeForce RTX 3090 GPU.
    \item \textbf{Recommendation Models}: All model parameters are initialized with a Gaussian distribution $\mathcal{N}(0, {0.01}^2)$.
    To obtain the optimal performance, we use grid search to tune the hyper-parameters. For model-specific hyper-parameters, we follow the suggestions from their original papers. Specifically, we set the learning rate to 0.001 and the embedding size to 32. The number of epochs is set to 20 for NMF and 200 for LightGCN.
    \item \textbf{Attacker}: For MLP, we set the L2 regularization weight to 1.0, the initial learning rate to $0.001$ and the maximal iteration to 500, leaving the other hyper-parameters at their defaults.
    For XGBoost, we use the xgboost package, setting the hyper-parameters as their default values. For RF, we set the n\_estimators to 100 and the max\_depth to 20. For AdaBoost, we set the n\_estimators to 50. For GBDT, we set the n\_estimators to 100. All these three models are implemented with scikit-learn 1.1.3~\footnote{https://scikit-learn.org/}.
    \item \textbf{Unlearning}: For the two-component loss, we set the learning rate to 1e-3. For ML-100K, ML-1M, LFM-2B and KuaiSAR, we investigate the hyper-parameter $\alpha$ to $\{2.5e^{-4}, 1.5e^{-6}, 5e^{-5}, 1e^{-5}\}$. The number of unlearning epochs is set to 500. For $\ell_r$, the value of $k$ is set to 20, while $\lambda$ and $\tau$ are set to 0.05 and 1e3 respectively. The $\lambda$ and $\tau$ are tuned using a grid search.


\end{itemize}
We run all models 10 times and report the average results.



\begin{table}
\caption{Results of unlearning performance (performance of attribute inference attack). The top results are highlighted in \textbf{bold}. 
InT-AU methods are represented in \texttt{typewriter font}.}
\label{tab:unlearn}
\centering
\resizebox{0.636\linewidth}{!}{
\begin{tabular}{ccc|cccc}
\toprule
\multirow{2}{*}{Dataset} &\multirow{2}{*}{Attribute} & \multirow{2}{*}{Method}  & \multicolumn{2}{|c|}{NMF} & \multicolumn{2}{c}{LightGCN} \\ 
& & & F1 & BAcc & \multicolumn{1}{|c}{F1} & BAcc\\
\midrule
\multirow{10}{*}{ML-100K} & \multirow{5}{*}{Gender} & Original & 0.6935 & 0.6870 & 0.6762 & 0.6784 \\
& & \texttt{Retrain}                               & 0.5037 & 0.5025 & \textbf{0.5195} & \textbf{0.5101} \\
& & \texttt{Adv-InT}                               & 0.5334 & 0.5673 & 0.5517 & 0.5401 \\
& & D2D-PR                                         & 0.5142 & 0.5074 & 0.5326 & 0.5219 \\
& & D2D-FR                                         & \textbf{0.4967} & \textbf{0.5016} & 0.5287 & 0.5113 \\

\cmidrule{2-7}
& \multirow{5}{*}{Age}           & Original & 0.6571 & 0.5335 & 0.6514 & 0.5179 \\
& & \texttt{Retrain}                        & 0.5653 & \textbf{0.3265} & 0.5715 & 0.3443 \\
& & \texttt{Adv-InT}                        & 0.5974 & 0.3761 & 0.6047 & 0.3688 \\
& & D2D-PR                                  & 0.5627 & 0.3342 & 0.5721 & 0.3446 \\
& & D2D-FR                                  & \textbf{0.5474} & 0.3321 & \textbf{0.5710} & \textbf{0.3443} \\

\midrule

\multirow{10}{*}{ML-1M} & \multirow{5}{*}{Gender}  & Original & 0.7602 & 0.7597 & 0.7204 & 0.7175 \\
& & \texttt{Retrain}                        & 0.5003 & 0.5009 & \textbf{0.5117} & \textbf{0.5056} \\
& & \texttt{Adv-InT}                        & 0.5574 & 0.5551 &  0.5874 & 0.5515 \\
& & D2D-PR                                  & 0.4979 & 0.5118 & 0.5229  & 0.5095 \\
& & D2D-FR                                  & \textbf{0.4944} & \textbf{0.5035} & 0.5187 & 0.5068 \\

\cmidrule{2-7}
& \multirow{5}{*}{Age}       & Original & 0.7166 & 0.6061 & 0.6994 & 0.5913 \\
& & \texttt{Retrain}                    & 0.5667 & 0.3338 & \textbf{0.5665} & \textbf{0.3334} \\
& & \texttt{Adv-InT}                    & 0.6125 & 0.3707 & 0.6114 & 0.3779 \\
& & D2D-PR                              & \textbf{0.5664} & 0.3334 & 0.5668 & 0.3341 \\
& & D2D-FR                              & 0.5665 & \textbf{0.3334} & 0.5671 & 0.3347 \\

\midrule 

\multirow{10}{*}{LFM-2B} & \multirow{5}{*}{Gender} & Original & 0.6836 & 0.6911 & 0.6679 & 0.6823 \\
& & \texttt{Retrain}                               & 0.5135 & \textbf{0.5062} & 0.5128 & 0.5065 \\
& & \texttt{Adv-InT}                               & 0.5547 & 0.5436 & 0.5643 & 0.5479 \\
& & D2D-PR                                         & 0.5139 & 0.5085 & 0.5145 & 0.5097 \\
& & D2D-FR                                         & \textbf{0.5121} & 0.5074 & \textbf{0.5114} & \textbf{0.5032} \\
\cmidrule{2-7}

& \multirow{5}{*}{Country} & Original & 0.5199 & 0.4257 & 0.5095 & 0.4187 \\
& & \texttt{Retrain}                         & 0.2214 & 0.1251 & 0.2215 & 0.1249 \\
& & \texttt{Adv-InT}                         & 0.2545 & 0.1434 & 0.2655 & 0.1572 \\
& & D2D-PR                                   & 0.2210 & \textbf{0.1248} & 0.2215 & 0.1255 \\
& & D2D-FR                                   & \textbf{0.2210} & 0.1249 & \textbf{0.2214} & \textbf{0.1247} \\
\midrule 

\multirow{10}{*}{KuaiSAR} & \multirow{5}{*}{Feat1}   & Original & 0.4433 & 0.2184  & 0.4525 & 0.2207  \\
& & \texttt{Retrain}                                & 0.3727 & 0.1427  & \textbf{0.3814} & \textbf{0.1413}  \\
& & \texttt{Adv-InT}                                & 0.4065 & 0.1608  & 0.4125 & 0.1681  \\
& & D2D-PR                                          & 0.3747 & 0.1429  & 0.3821 & 0.1427  \\
& & D2D-FR                                          & \textbf{0.3713} & \textbf{0.1427}  & 0.3819 & 0.1426  \\
\cmidrule{2-7}

& \multirow{5}{*}{Feat2} & Original & 0.8261 & 0.8242 & 0.8065 & 0.7973 \\
& & \texttt{Retrain}                        & 0.5565 & 0.5603 & 0.5556 & \textbf{0.5471} \\
& & \texttt{Adv-InT}                        & 0.6107 & 0.5985 & 0.5957 & 0.5821 \\
& & D2D-PR                                  & 0.5638 & 0.5600 & 0.5574 & 0.5495 \\
& & D2D-FR                                  & \textbf{0.5534} & \textbf{0.5587} & \textbf{0.5543} & 0.5476 \\
\bottomrule
\end{tabular}
}
\end{table}

\begin{table*}
\caption{Results of recommendation performance. The top results are highlighted in \textbf{bold} (except for Retrain). InT-AU methods are represented in \texttt{typewriter font}. The top results of InT-AU methods are \underline{underlined}.}
\label{tab:rec}
\centering
\resizebox{1.0\linewidth}{!}{  
\begin{tabular}{ccc|cccc|cccc}
\toprule
\multirow{2}{*}{Dataset} & \multirow{2}{*}{Attribute} & \multirow{2}{*}{Method} & \multicolumn{4}{c}{NMF} & \multicolumn{4}{c}{LightGCN} \\

& & & NDCG@5 & HR@5 & NDCG@10 & HR@10 & NDCG@5 & HR@5 & NDCG@10 & HR@10\\
\midrule
\multirow{10}{*}{ML-100k} & \multirow{5}{*}{Gender} & Original  & 0.0649 & 0.1007 & 0.0835 & 0.1601 & 0.0668 & 0.1043 & 0.0859 & 0.1663\\
& & \texttt{Retrain}                                            & 0.0646 & 0.1007 & \textbf{0.0834} & \textbf{0.1603} & \textbf{0.0667} & \textbf{0.1045} & \textbf{0.0855} & \textbf{0.1662} \\
& & \texttt{Adv-InT}                                            & 0.0623 & 0.0965 & 0.0799 & 0.1523 & 0.0644 & 0.1006 & 0.0812 & 0.1524\\
& & D2D-PR                                                      & 0.0645 & 0.0997 & 0.0807 & 0.1506 & 0.0657 & 0.1034 & 0.0838 & 0.1597\\
& & D2D-FR                                                      & \underline{\textbf{0.0649}} & \underline{\textbf{0.1008}} & \underline{0.0832} & \underline{0.1591} & \underline{0.0665} & \underline{0.1043} & \underline{0.0854} & \underline{0.1659}\\
\cmidrule{2-11}
& \multirow{5}{*}{Age}     & Original & 0.0649 & 0.1007 & 0.0835 & 0.1601 & 0.0668 & 0.1043 & 0.0859 & 0.1663 \\
& & Retrain                                                     & \textbf{0.0644} & \textbf{0.1002} & 0.0807 & \textbf{0.1531} & 0.0649 & \textbf{0.1021} & 0.0841 & 0.1574 \\
& & Adv-InT                                                     & 0.0605 & 0.0941 & 0.0782 & 0.1497 & 0.0625 & 0.0975 & 0.0792 & 0.1556 \\
& & D2D-PR                                                      & 0.0617 & 0.0954 & 0.0789 & 0.1485 & 0.0624 & 0.0983 & 0.0789 & 0.1545 \\
& & D2D-FR                                                      & \underline{0.0642} & \underline{0.0997} & \underline{\textbf{0.0810}} & \underline{0.1527} & \underline{\textbf{0.0651}} & \underline{0.1006} & \underline{\textbf{0.0845}} & \underline{\textbf{0.1581}} \\
\midrule

\multirow{10}{*}{ML-1M} & \multirow{5}{*}{Gender}   & Original                           & 0.0432 & 0.0679 & 0.0574 & 0.1121  & 0.0422 & 0.0664 & 0.0562 & 0.1097\\
& & Retrain                                                     & 0.0431 & 0.0675 & 0.0562 & \textbf{0.1108}  & 0.0421 & \textbf{0.0665} & 0.0557 & \textbf{0.1088} \\
& & Adv-InT                                                     & 0.0408 & 0.0651 & 0.0546 & 0.1062  & 0.0397 & 0.0634 & 0.0532 & 0.1035 \\
& & D2D-PR                                                      & 0.0414 & 0.0654 & 0.0543 & 0.1053  & 0.0405 & 0.0651 & 0.0546 & 0.1042 \\
& & D2D-FR                                                      & \underline{\textbf{0.0433}} & \underline{\textbf{0.0681}} & \underline{\textbf{0.0568}} & \underline{0.1104}  & \underline{\textbf{0.0421}} & \underline{0.0664} & \underline{\textbf{0.0559}} & \underline{0.1087} \\
\cmidrule{2-11}

&\multirow{5}{*}{Age}   & Original                & 0.0432 & 0.0679 & 0.0574 & 0.1121 & 0.0422 & 0.0664 & 0.0562 & 0.1097 \\
& & Retrain                                         & \textbf{0.0433} & 0.0678 & \textbf{0.0566} & \textbf{0.1092} & \textbf{0.0423} & 0.0662 & 0.0555 & \textbf{0.1081} \\
& & Adv-InT                                         & 0.0386 & 0.0626 & 0.0527 & 0.1064 & 0.0382 & 0.0621 & 0.0528 & 0.1058 \\
& & D2D-PR                                          & 0.0403 & 0.0647 & 0.0542 & 0.1078 & 0.0405 & 0.0645 & 0.0533 & 0.1056 \\
& & D2D-FR                                          & \underline{0.0432} & \underline{\textbf{0.0684}} & \underline{0.0561} & \underline{0.1087} & \underline{0.0422} & \underline{\textbf{0.0669}} & \underline{\textbf{0.0556}} & \underline{0.1077} \\
\midrule

\multirow{10}{*}{LFM-2B} & \multirow{5}{*}{Gender} & Original                            & 0.0089 & 0.0151 & 0.0123 & 0.0258 & 0.0104 & 0.0176 & 0.0141 & 0.0273\\
& & Retrain                                                     & 0.0088 & 0.0149 & 0.0124 & \textbf{0.0261} & 0.0102 & \textbf{0.0177} & 0.0139 & 0.0270\\
& & Adv-InT                                                     & 0.0086 & 0.0143 & 0.0119 & 0.0252 & 0.0098 & 0.0165 & 0.0135 & 0.0265\\
& & D2D-PR                                                      & 0.0088 & 0.0145 & \underline{\textbf{0.0124}} & 0.0256 & 0.0097 & 0.0168 & 0.0137 & 0.0264\\
& & D2D-FR                                                      & \underline{\textbf{0.0089}} & \underline{\textbf{0.0151}} & 0.0123 & \underline{0.0260} & \underline{\textbf{0.0102}} & \underline{0.0173} & \underline{\textbf{0.0143}} & \underline{\textbf{0.0271}}\\
\cmidrule{2-11}
& \multirow{5}{*}{Country}   & Original    & 0.0089 & 0.0151 & 0.0123 & 0.0258 & 0.0104 & 0.0176 & 0.0141 & 0.0273\\
& & Retrain                                                        & \textbf{0.0086} & \textbf{0.0145} & 0.0112 & \textbf{0.0234} & \textbf{0.0104} & \textbf{0.0165} & 0.0135 & 0.0253 \\
& & Adv-InT                                                        & 0.0083 & 0.0139 & 0.0109 & 0.0230 & 0.0097 & 0.0159 & 0.0130 & 0.0251 \\
& & D2D-PR                                                         & 0.0080 & 0.0135 & 0.0110 & 0.0230 & 0.0098 & 0.0161 & 0.0132 & 0.0249 \\
& & D2D-FR                                                         & \underline{0.0085} & \underline{0.0140} & \underline{\textbf{0.0114}} & \underline{0.0231} & \underline{0.0101} & \underline{0.0164} & \underline{\textbf{0.0135}} & \underline{\textbf{0.0255}} \\

\midrule
\multirow{10}{*}{KuaiSAR} & \multirow{5}{*}{Feat1}  & Original    & 0.0118 & 0.0186 & 0.0160 & 0.0318 & 0.0131 & 0.0197 & 0.0175 & 0.0334  \\
& & Retrain                                                       & 0.0114 & \textbf{0.0184} & \textbf{0.0152} & 0.0309 & \textbf{0.0128} & \textbf{0.0193} & 0.0171 & 0.0327 \\
& & Adv-InT                                                       & 0.0112 & 0.0175 & 0.0149 & 0.0303 & 0.0124 & 0.0186 & 0.0165 & 0.0317 \\
& & D2D-PR                                                        & 0.0111 & 0.0177 & \underline{0.0151} & 0.0301 & 0.0125 & 0.0185 & 0.0167 & 0.0318 \\
& & D2D-FR                                                        & \underline{\textbf{0.0115}} & \underline{0.0183} & 0.0150 & \underline{\textbf{0.0310}} & \underline{0.0127} & \underline{0.0193} & \underline{\textbf{0.0173}} & \underline{\textbf{0.0328}} \\
\cmidrule{2-11}
& \multirow{5}{*}{Feat2} & Original    & 0.0118 & 0.0186 & 0.0160 & 0.0318 & 0.0131 & 0.0197 & 0.0175 & 0.0334 \\
& & Retrain                                                       & 0.0115 & \textbf{0.0179} & \textbf{0.0156} & 0.0316 & \textbf{0.0129} & \textbf{0.0188} & 0.0168 & \textbf{0.0332} \\
& & Adv-InT                                                       & 0.0109 & 0.0171 & 0.0151 & 0.0304 & 0.0124 & 0.0185 & 0.0164 & 0.0324 \\
& & D2D-PR                                                        & 0.0113 & 0.0173 & 0.0153 & 0.0306 & 0.0122 & 0.0184 & 0.0165 & 0.0323 \\
& & D2D-FR                                                        & \underline{\textbf{0.0116}} & \underline{0.0176} & \underline{0.0154} & \underline{\textbf{0.0316}} & \underline{0.0125} & \underline{0.0186} & \underline{\textbf{0.0168}} & \underline{0.0331} \\
\bottomrule
\end{tabular}
}
\end{table*}

\begin{figure*}[t]
\centering
\subfigure[\textbf{ML-100K (original)}]{\includegraphics[width=3.4cm]{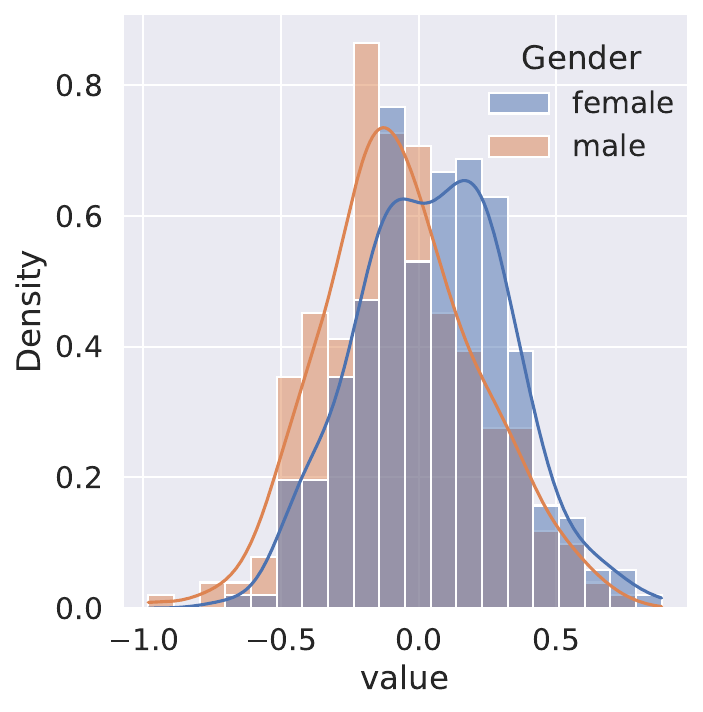}}
\subfigure[\textbf{ML-1M (original)}]{\includegraphics[width=3.4cm]{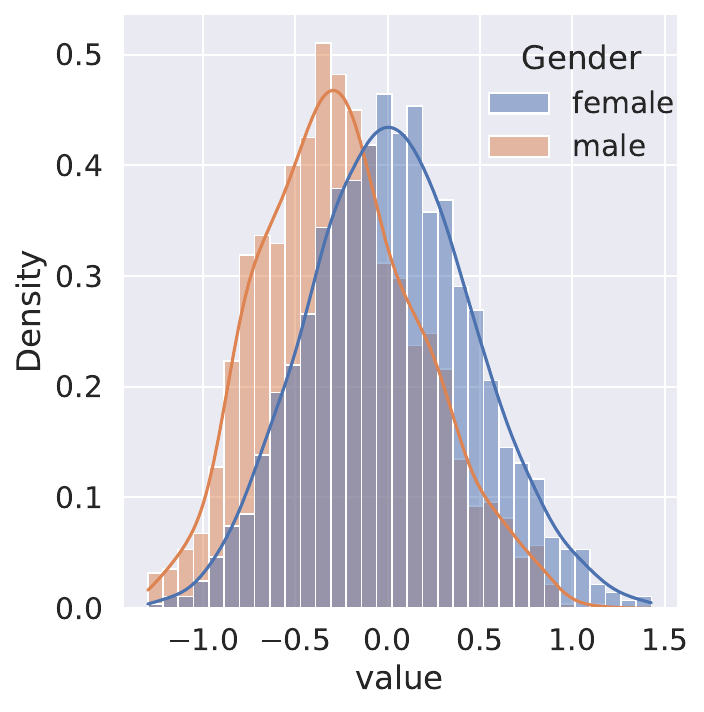}}
\subfigure[\textbf{LFM-2B (original)}]{\includegraphics[width=3.4cm]{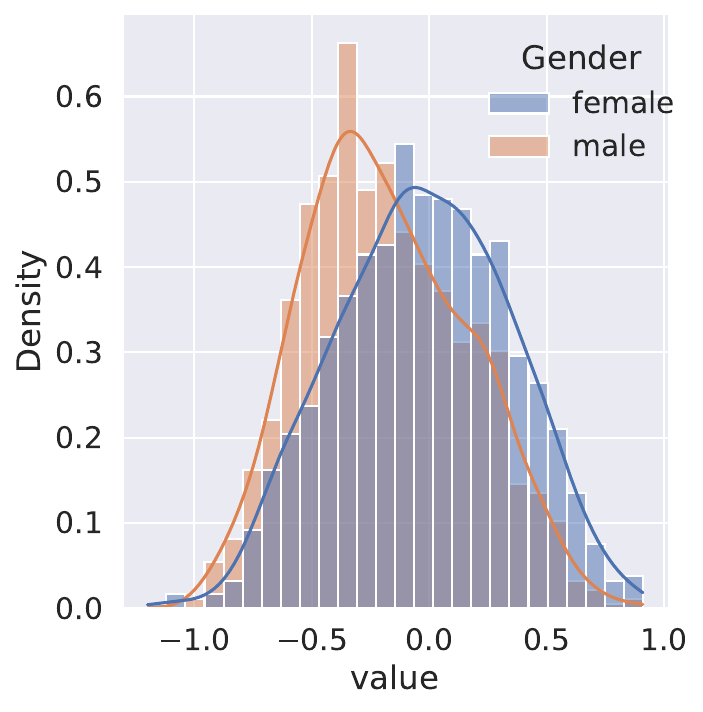}}
\subfigure[\textbf{KuaiSAR (original)}]{\includegraphics[width=3.4cm]{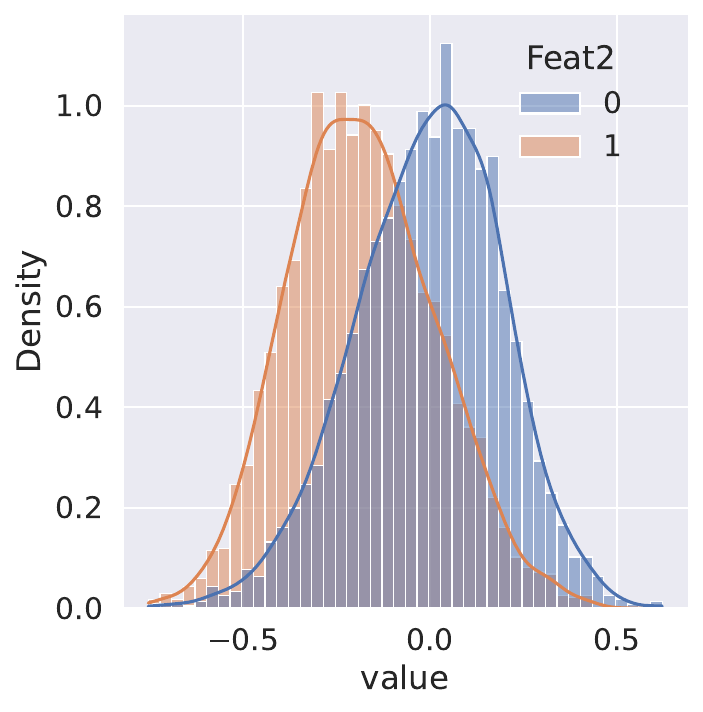}}
\subfigure[\textbf{ML-100K (ours)}]{\includegraphics[width=3.4cm]{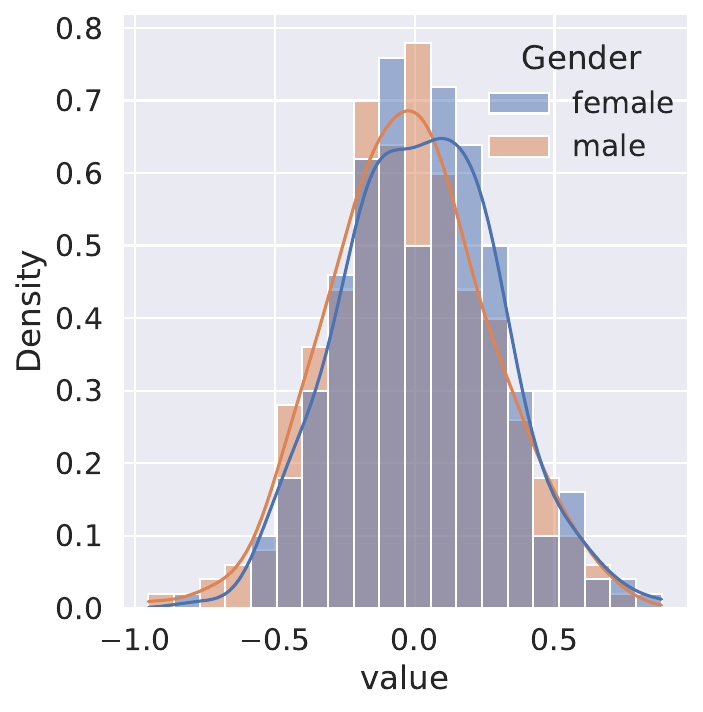}}
\subfigure[\textbf{ML-1M (ours)}]{\includegraphics[width=3.4cm]{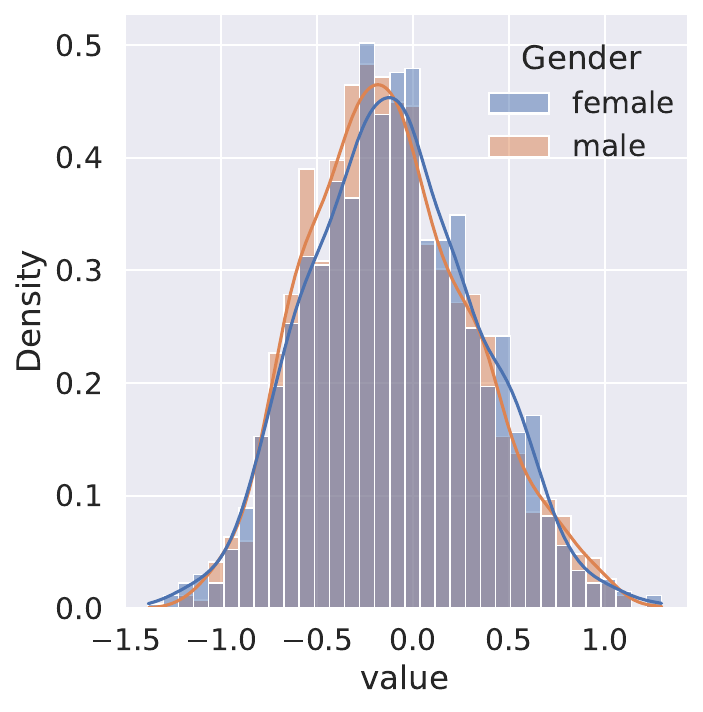}}
\subfigure[\textbf{LFM-2B (ours)}]{\includegraphics[width=3.4cm]{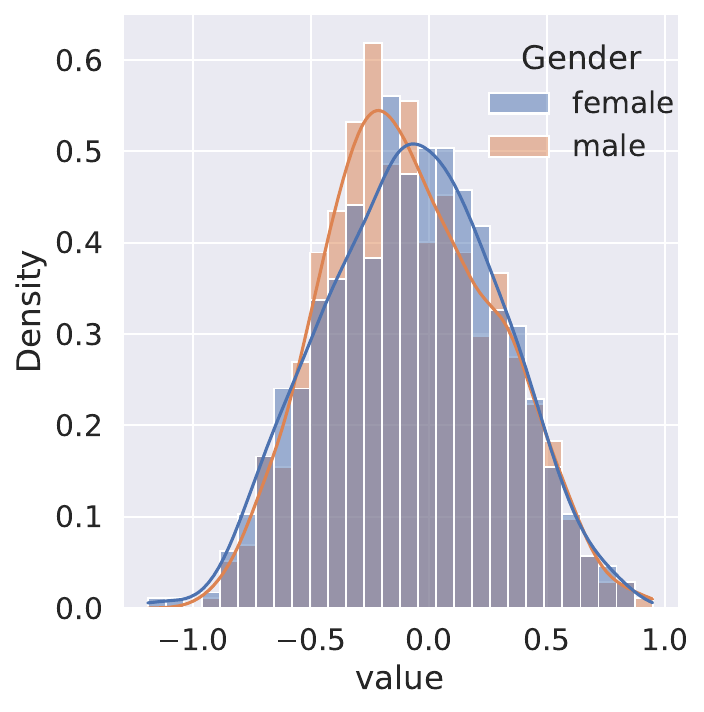}}
\subfigure[\textbf{KuaiSAR (ours)}]{\includegraphics[width=3.4cm]{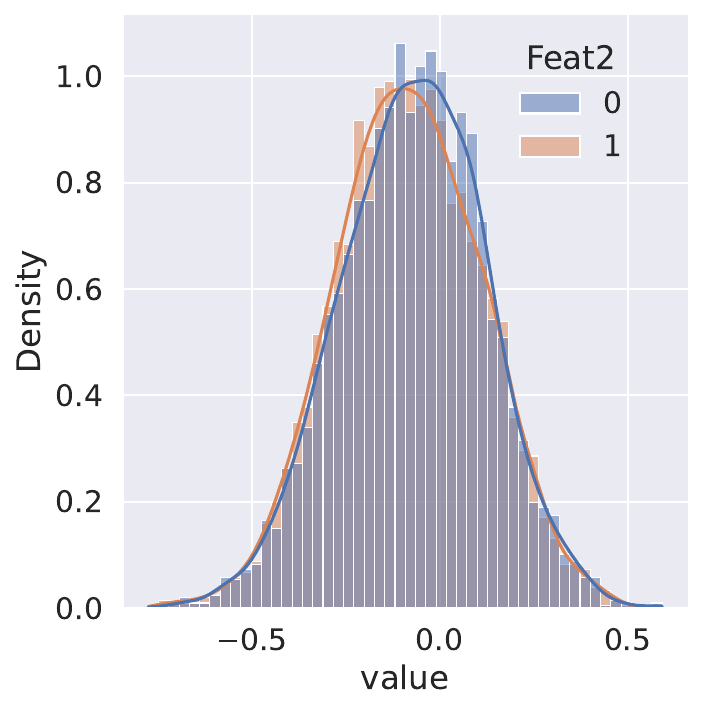}}
\caption{Distribution of user embedding in the first dimension on NMF}
\label{fig:emb_distribution}
\end{figure*}

\begin{table}
\caption{Running time of unlearning methods.}
\centering
\resizebox{0.65\linewidth}{!}{
\label{tab:time}
\begin{tabular}{cc|cccc}
\toprule
\multicolumn{2}{c}{Time (s)} & Retrain & Adv-InT & D2D-PR & D2D-FR\\
\midrule
\multirow{2}{*}{ML-100K (Age)} & NMF  & 85.43 & 159.75 & 5.46 & 4.76\\
& LightGCN  & 229.77 & 415.45 & 13.31 & 11.57\\
\midrule
\multirow{2}{*}{ML-1M (Age)} & NMF  & 943.57 & 1266.24 & 78.21 & 72.66\\
& LightGCN  & 1839.73 & 2414.52 & 167.85 & 143.44\\
\midrule
\multirow{2}{*}{LFM-2B (Country)} & NMF  & 1148.52 & 1457.82 & 95.51 & 47.92\\
& LightGCN  & 2264.55 & 2617.21 & 193.64 & 92.35\\
\midrule
\multirow{2}{*}{KuaiSAR (Feat1)} & NMF  & 971.23 & 1344.35 & 97.53 & 37.92\\
& LightGCN  & 1874.53 & 2506.38 & 179.24 & 76.51\\
\bottomrule
\end{tabular}
}
\end{table}

\subsection{Results and Discussions}
%

\subsubsection{Unlearning Performance (RQ1) \label{exp:rq1}}
Unlearning the target attribute is the primary goal of PoT-AU.
The performance of unlearning is evaluated by the performance of attacker, i.e., MLP.
We train the attacker on training set, and report its performance on the testing set. 
To comprehensively evaluate attacking performance, we report two metrics, including F1 score and BAcc, in Table~\ref{tab:unlearn}.
%
%
%
We have the following observations from the above results. 
Firstly, attackers achieve an average F1 Score of 0.66 and BAcc of 0.59 on the original embedding, indicating that information on the user's attribute in user embeddings can be released to attackers.
Secondly, all methods can unlearn attribute information contained in user embeddings to varying degrees.
Retrain, Adv-InT, D2D-PR and D2D-FR decrease the F1 Score by 27.33\%, 20.79\%, 26.93\%, and 27.55\%, respectively, on average. Meanwhile, D2D-PR, D2D-FR and Retrain can decrease the BAcc by 37.23\%, 37.6\% and 37.72\% on average. In comparison, Adv-InT can only decrease the BAcc by 30.97\%. For binary attributes, e.g., gender, the BAcc of attacker after unlearning with D2D-FR method is equivalent to that of a random attacker, which indicates that our proposed D2D-FR can effectively unlearn the private information of recommendation models.
Thirdly, as shown in Table~\ref{tab:unlearn}, although without the access to training data, our D2D-based methods demonstrate comparable unlearning performance with Retrain in general.
%



\nosection{Summary} Compared with Adv-InT, D2D-PR and D2D-FR is more effective in unlearning, which protects the user's attributes by making them indistinguishable to the attacker.

\subsubsection{Recommendation Performance (RQ2) \label{exp:rq2}}
Recommendation performance is the other important goal in the PoT-AU problem, since attribute unlearning is usually at the expense of model accuracy. To answer RQ2, we use NDCG and HR to evaluate recommendation performance after unlearning and truncate the ranked list at 5 and 10 for both metrics.
As shown in Table~\ref{tab:rec}, unlearning methods also affect recommendation performance.
Compared with the original performance, Adv-InT and D2D-PR decrease the NDCG by 6.25\% and 4.88\%, and decrease the HR by 5.81\% and 5.05\%, respectively, on average. However, D2D-FR only has an average degradation of 1.91\% on NDCG and 2.14\% on HR. Retrain has an average degradation of 1.79\% on NDCG and 2.05\% on HR, which is slightly better than D2D-FR.
Interestingly, 
%
D2D-FR, which is devised to make attributes indistinguishable, could accidentally diminish the negative discrimination to enhance recommendation performance. As shown in Fig.~\ref{fig:emb_distribution}, the embeddings of users with different attribute categories after unlearning are indistinguishable.

\nosection{Summary} Compared to Adv-InT and D2D-PR, D2D-FR preserves the recommendation performance to a greater extent while achieving the objective of unlearning, approaching the level of Retrain.

\subsubsection{Efficiency (RQ3) \label{exp:rq3}}
To answer RQ3, we use running time to evaluate the efficiency of unlearning methods. Note that Age, Country and Feat1 are chosen as the targets for unlearning in this context.
From Table~\ref{tab:time}, we observe that i) our proposed PoT-AU methods (D2D-PR and D2D-FR) significantly outperform InT-AU methods (Retrain and Adv-InT).
This is because PoT-AU methods can be viewed as a fine-tuning process on an existing model, providing them with inherent efficiency compared to InT-AU methods; ii) By incorporating our proposed distinguishability loss to the original recommendation loss and retraining from scratch, Retrain outperforms Adv-InT. As a baseline method, Retrain provides a new path for InT-AU methods to explore; iii) In the scenario of multi-class attribute unlearning, D2D-FR is more efficient than D2D-PR. Compared to D2D-PR, D2D-FR reduces the running time by 51.48\% and 58.66\% on LFM-2B and KuaiSAR respectively. By adopting the $\ell_u$ which introduces an \textit{anchor distribution} to compute distance, D2D-FR can effectively reduce the computational complexity of unlearning.

\subsubsection{Parameter Sensitivity (RQ4) \label{exp:rq4}}
To answer RQ4, we investigate the performance fluctuations of our method with varied hyper-parameters, i.e., the trade-off coefficient $\alpha$ and the length of rank list $k$ for $\ell_r$. 
Specifically, we tune the value of $\alpha$ and $k$ while keeping the other hyper-parameters unchanged. 

\begin{figure*}[t]
\centering
\includegraphics[width=13cm]{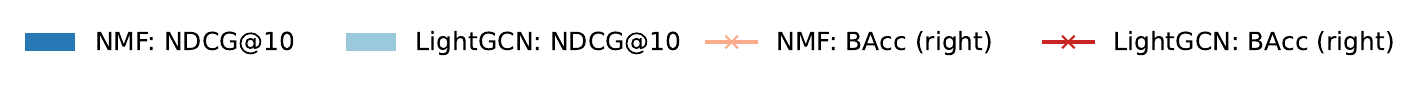}\\
\subfigure[\textbf{ML-100K (Gender)}]{\includegraphics[width=6cm]{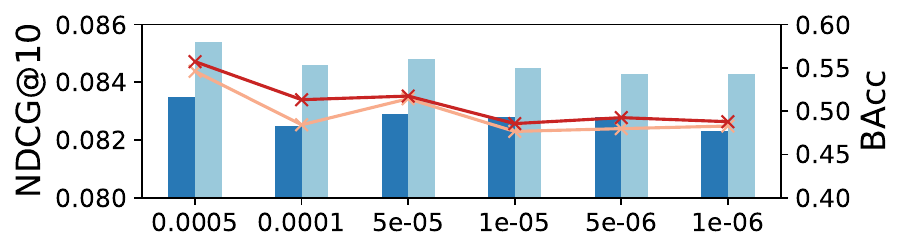} \label{fig:ml-100k_alpha}}
\subfigure[\textbf{ML-1M (Gender)}]{\includegraphics[width=6cm]{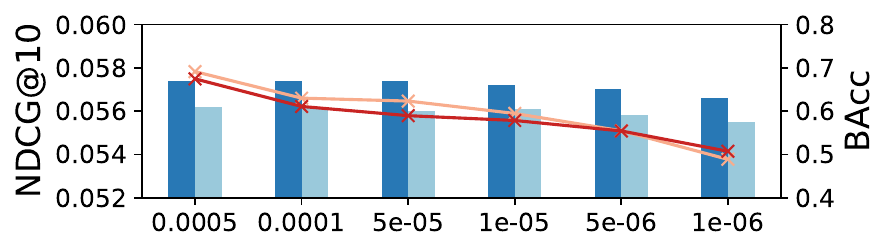} \label{fig:ml-1m_alpha}}
\subfigure[\textbf{LFM-2B (Gender)}]{\includegraphics[width=6cm]{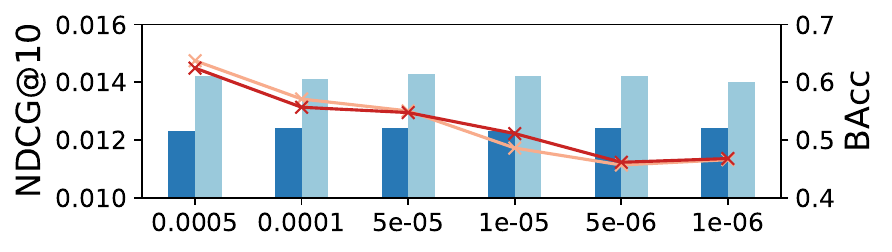} \label{fig:lfm-2b_alpha}}
\subfigure[\textbf{KuaiSAR (Feat2)}]{\includegraphics[width=6cm]{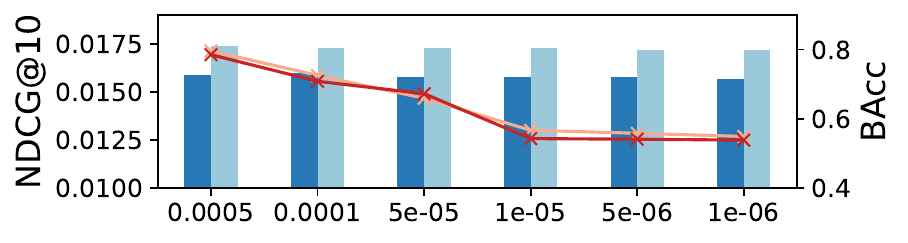} \label{fig:kuaisar_alpha}}
\caption{Effect of the hyper-parameter $\alpha$.}
\label{fig:para_alpha}
\end{figure*}

\begin{figure*}[t]
\centering
\subfigure[\textbf{ML-100K (Rec)}]{\includegraphics[width=3.4cm]{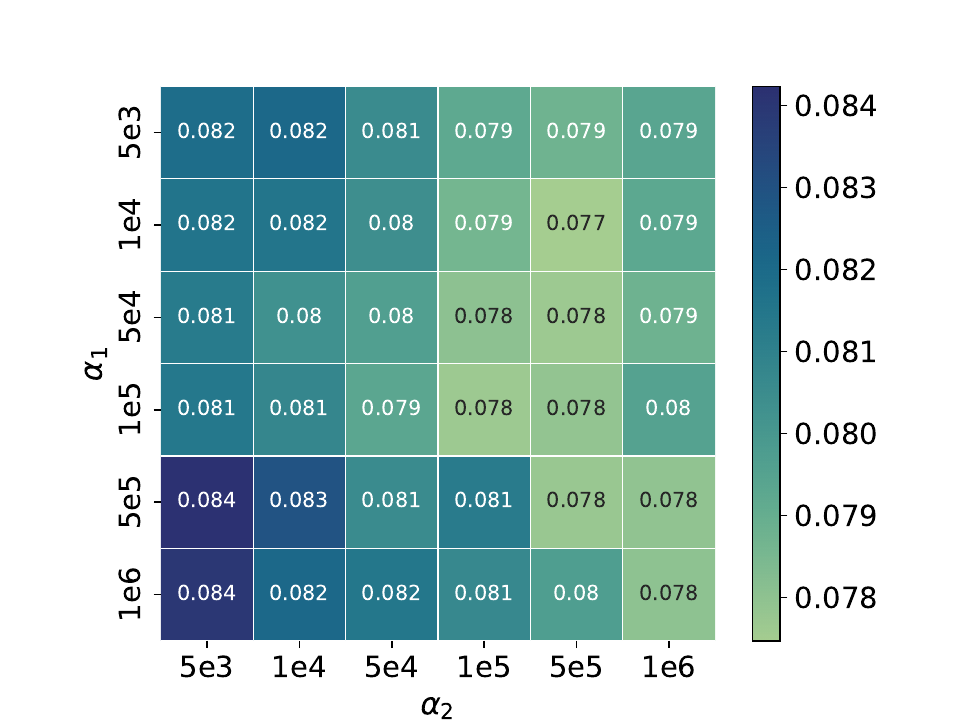}}
\subfigure[\textbf{ML-1M (Rec)}]{\includegraphics[width=3.4cm]{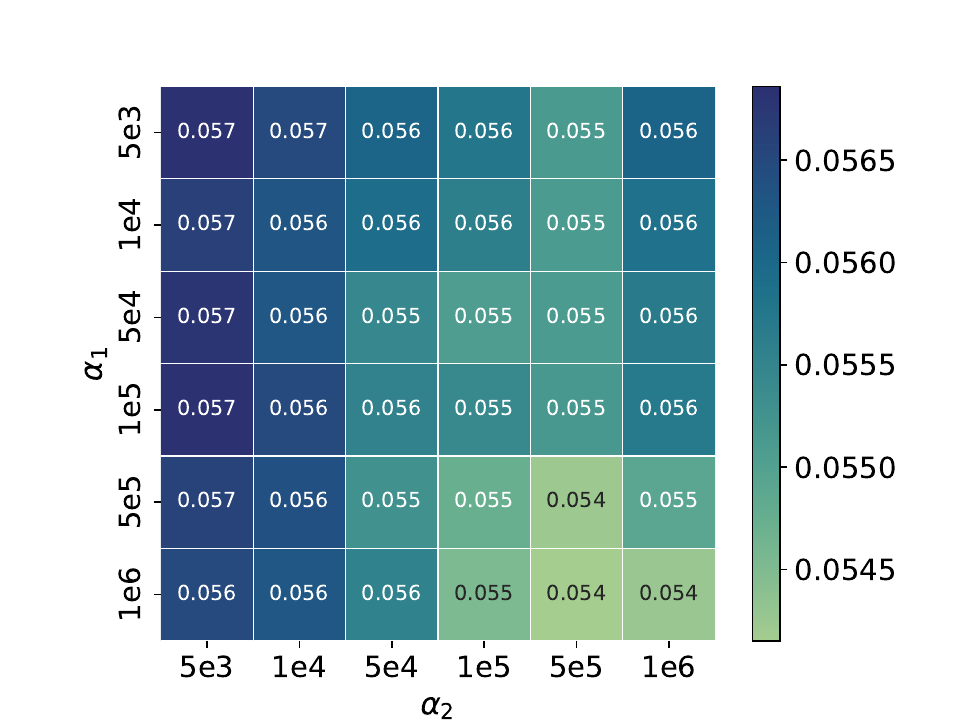}}
\subfigure[\textbf{LFM-2B (Rec)}]{\includegraphics[width=3.4cm]{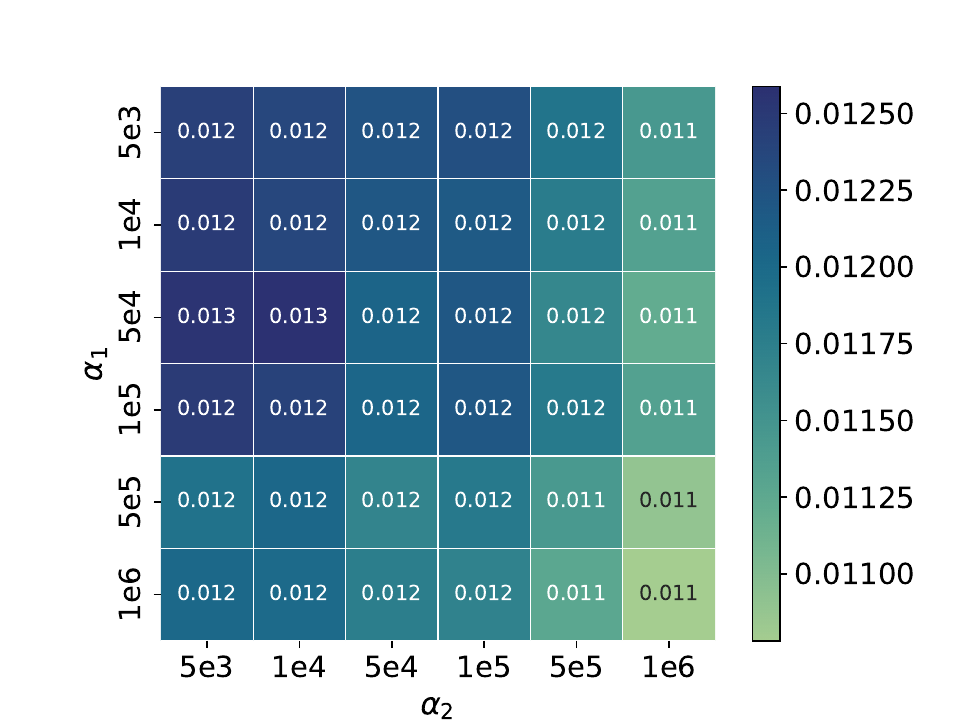}}
\subfigure[\textbf{KuaiSAR (Rec)}]{\includegraphics[width=3.4cm]{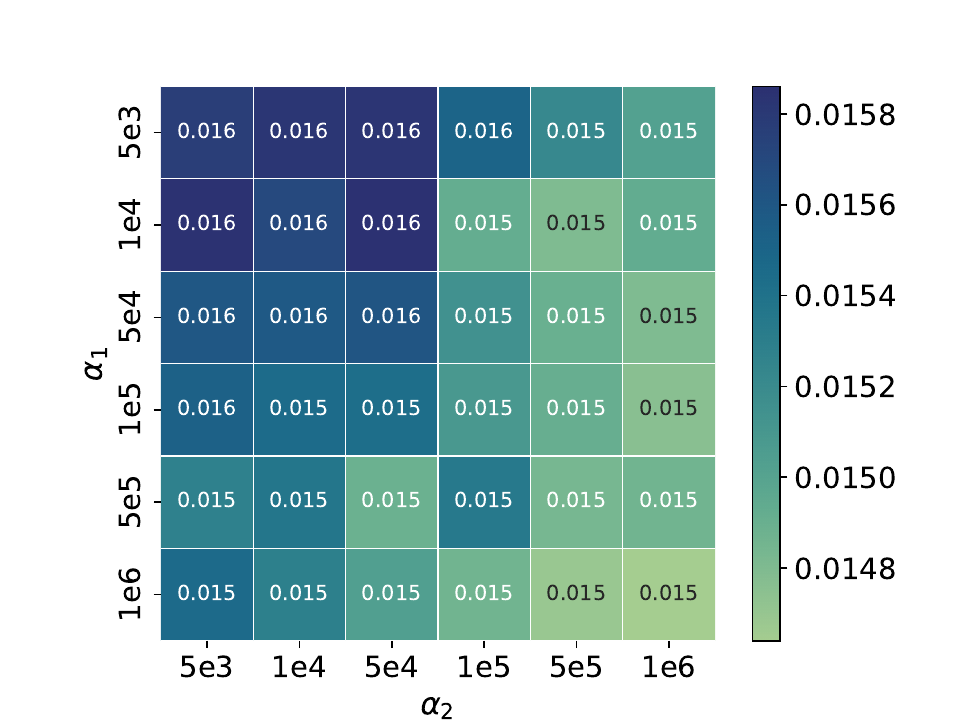}}
\subfigure[\textbf{ML-100K (Unlearn)}]{\includegraphics[width=3.4cm]{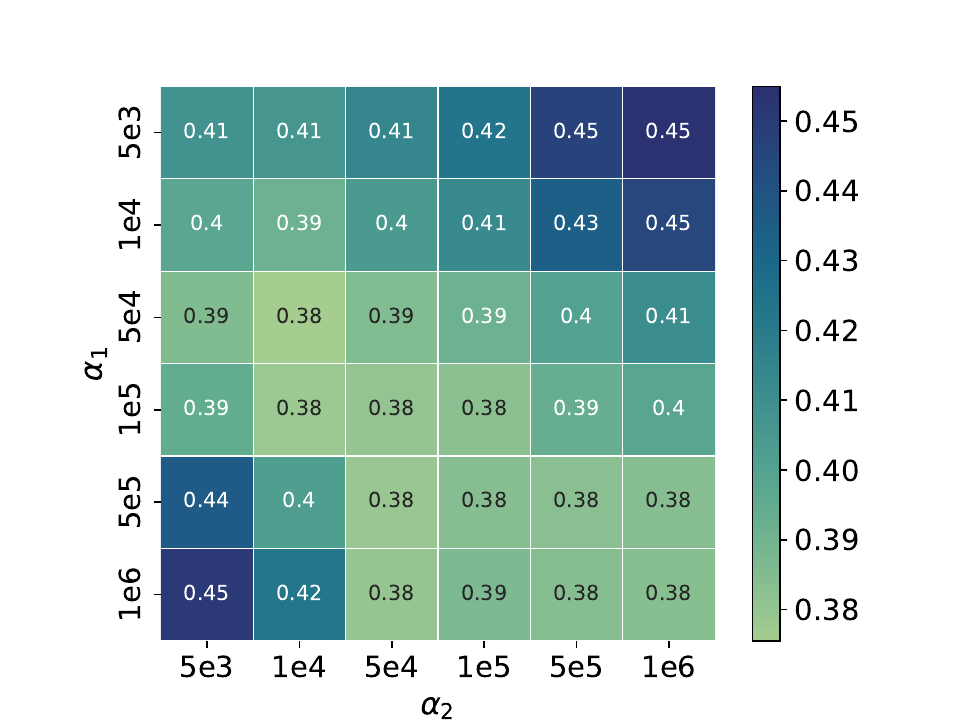}}
\subfigure[\textbf{ML-1M (Unlearn)}]{\includegraphics[width=3.4cm]{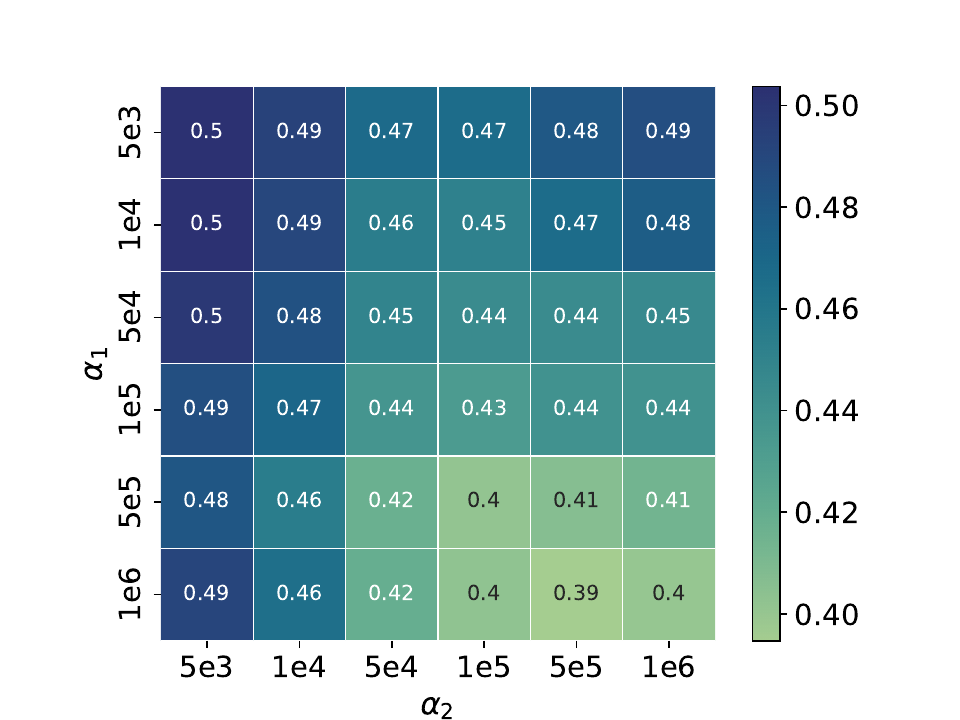}}
\subfigure[\textbf{LFM-2B (Unlearn)}]{\includegraphics[width=3.4cm]{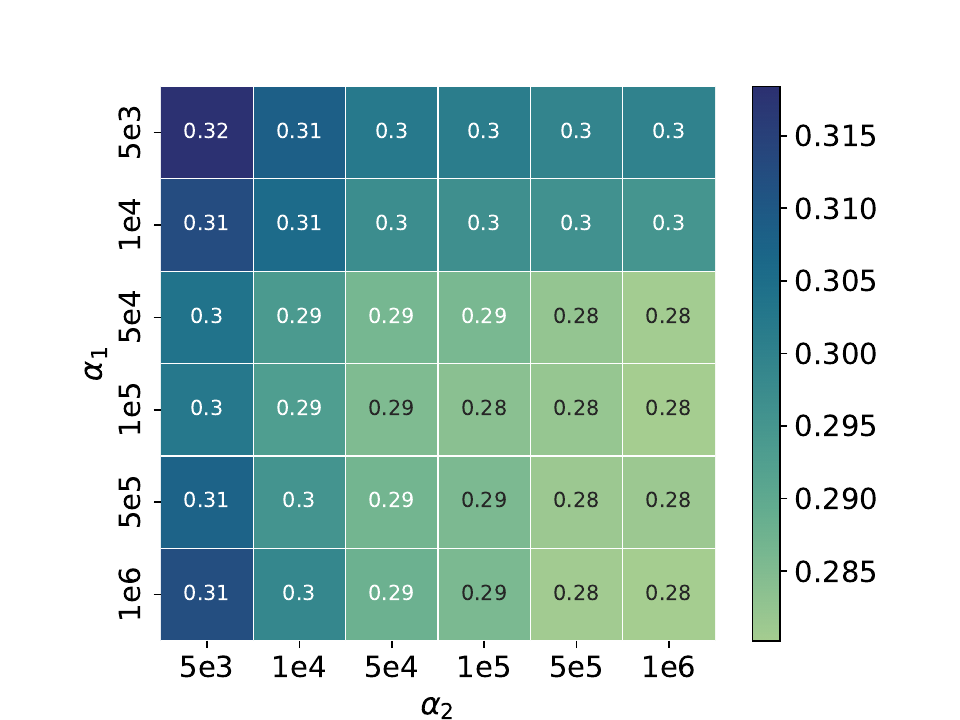}}
\subfigure[\textbf{KuaiSAR (Unlearn)}]{\includegraphics[width=3.4cm]{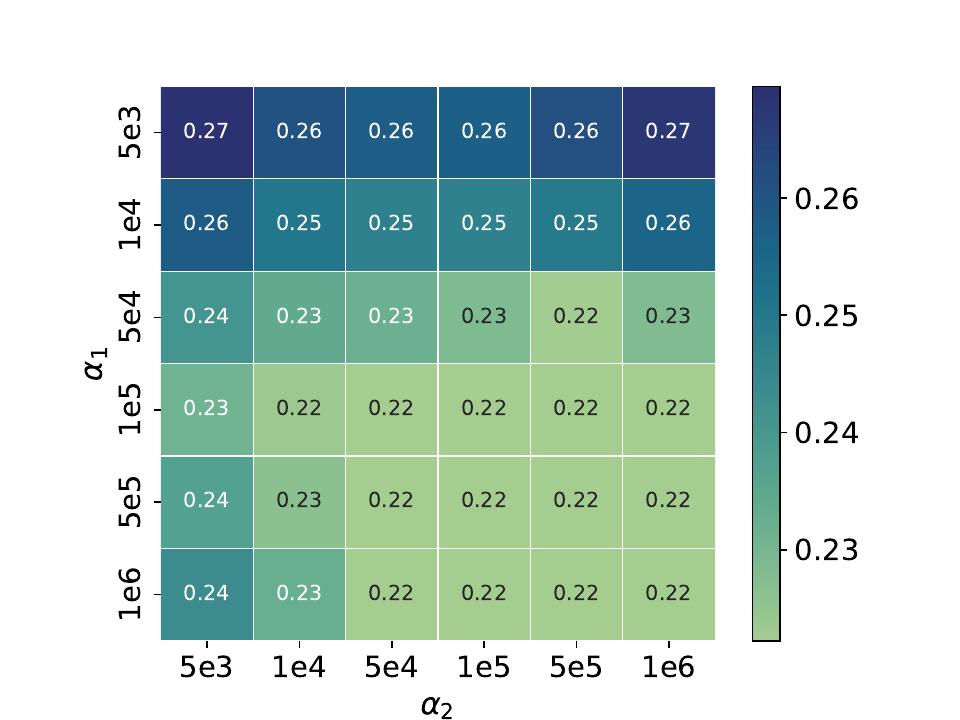}}
\caption{Effect of the hyper-parameter $\alpha_{1}$ and $\alpha_{2}$ in the scenario of unlearning multiple attributes on NMF. The first and second lines represent NDCG@10 (recommendation performance) and wBAcc (unlearning performance) respectively.}
\label{fig:hyper_para_multi}
\end{figure*}

\begin{itemize}[leftmargin=*] \setlength{\itemsep}{-\itemsep}
    \item \textbf{Trade-off parameter $\alpha$}. As shown in Fig.~\ref{fig:para_alpha}, we use BAcc and NDCG@10 to represent the performance of unlearning and recommendation respectively. We observe that the NDCG@10 of our proposed method, i.e., D2D-FR, is robust with different $\alpha$. Meanwhile, reducing the value of $\alpha$ results in a decrease in BAcc. The above observations indicate that D2D-FR can enhance unlearning performance with insignificant performance degradation for recommendation.
    \item \textbf{Trade-off parameter for unlearning multiple attributes $\alpha_1$ and $\alpha_2$}. In practice, simultaneous unlearning of multiple attributes unfolds naturally. We also build a loss function under our proposed two-component framework to probe this scenario. Specifically, it computes as
    \begin{equation}
        L(\boldsymbol{\theta}) = \ell_r + \alpha_1 \ell_{u1} + \alpha_2 \ell_{u2},
    \end{equation}
    where $\ell_{u1}$ and $\ell_{u2}$ denote the first and second attributes respectively.
    We use NDCG@10 to evaluate recommendation performance. As there are two attributes, we build a weighted-BAcc to comprehensively evaluate unlearning performance. Specifically, it computes as
    \begin{equation}
    \text{wBAcc} = \frac{\sum_{i=1}^{T}c_{i}*BAcc_{i}}{\sum_{i=1}^{T}c_{i}},
    \end{equation}
    where $c_{i}$ denotes the label number of $i$-th attribute, the $BAcc_{i}$ denotes the $BAcc$ of AIA regarding to $i$-th attribute.
    As shown in Fig.~\ref{fig:hyper_para_multi}, we observe a trade-off between the performance of unlearning and recommendation, consistent with the scenario of unlearning a single attribute. However, the fluctuation of different hyper-parameters is insignificant, indicating the robustness of our proposed method. Selecting apt trade-off parameters appears straightforward, with our chosen values for ($\alpha_1$, $\alpha_2$) are (1e4, 5e3), (1e5, 5e4), (1e4, 5e3), and (5e4, 1e4) for ML-100K, ML-1M, LFM-2B, and KuaiSAR respectively.
    In addition, we report the performance w.r.t. recommendation unlearning of our chosen trade-off parameter in Table~\ref{tab:result_multiple}. 
    It is evident that our proposed method can significantly reduce the accuracy of the attacker, effectively unlearning the target attribute. At the same time, our method has a limited negative impact on recommendation performance, and in some cases, it even results in an increase, i.e., LFM-2B.

    \item \textbf{Length of rank list} $k$. The $k$ in $\ell_r$ represents the length of recommended item list for alignment. As shown in Table~\ref{tab:para_k}, D2D-FR with different $k$ can achieve the same unlearning effectiveness. However, larger or smaller $k$ both can reduce the recommendation effectiveness. Specifically, a smaller $k$ cannot retain the preference information in top-$k$ recommended item list, as $k$ increases, the top-$k$ ranked items may contain more noise. In our experiments, we set the $k$ to 20 for optimal performance of recommendation.
\end{itemize}

%
%
%

\subsubsection{Analysis of $\ell_r$ (RQ5) \label{exp:rq5}}

\begin{table*}
\caption{Performance w.r.t. recommendation and unlearning in the scenario of unlearning multiple attributes on NMF, where the change (\%) refers to the change of values before and after unlearning.}
\label{tab:result_multiple}
\centering
\resizebox{0.55\linewidth}{!}{  
\begin{tabular}{c|cc|cc}
\toprule
\multirow{2}{*}{Dataset} & \multicolumn{2}{c|}{NDCG@10} & \multicolumn{2}{c}{wBAcc} \\
& Value & Change (\%) & Value & Change(\%)\\
\midrule
ML-100k & 0.0816 & -2.28 & 0.4029 & -32.27 \\
ML-1M & 0.0556 & -3.14 & 0.4394 & -34.17 \\
LFM-2B & 0.0125 & 1.63 & 0.3053 & -36.24 \\
KuaiSAR & 0.0156 & -2.50 & 0.2337 & -33.80 \\
\bottomrule
\end{tabular}
}
\end{table*}

\begin{table}
\caption{Effect of the hyper-parameter $k$ on ML-1M.}
\centering
\resizebox{0.55\linewidth}{!}{ 
\label{tab:para_k}
\begin{tabular}{cc|cccc}
\toprule
Models & $k$ & F1 & BAcc & NDCG@10 & HR@10 \\
\midrule
\multirow{4}{*}{NMF} & 10  & 0.5664 & 0.3333 & 0.0552 & 0.1068\\
& 20  & 0.5665 & 0.3334 & 0.0561 & 0.1087\\
& 30  & 0.5665 & 0.3335 & 0.0553 & 0.1084\\
& 50  & 0.5664 & 0.3333 & 0.0541 & 0.1071\\
\midrule
\multirow{4}{*}{LightGCN} & 10  & 0.5669 & 0.3342 & 0.0535 & 0.1064\\
& 20  & 0.5671 & 0.3341 & 0.0548 & 0.1073\\
& 30  & 0.5673 & 0.3343 & 0.0542 & 0.1069\\
& 50  & 0.5673 & 0.3343 & 0.0537 & 0.1062\\
\bottomrule
\end{tabular}
}
\end{table}

\begin{figure}[t]
\centering
\subfigure[\textbf{HR@10 on ML1M}]{\includegraphics[width=4.3cm]{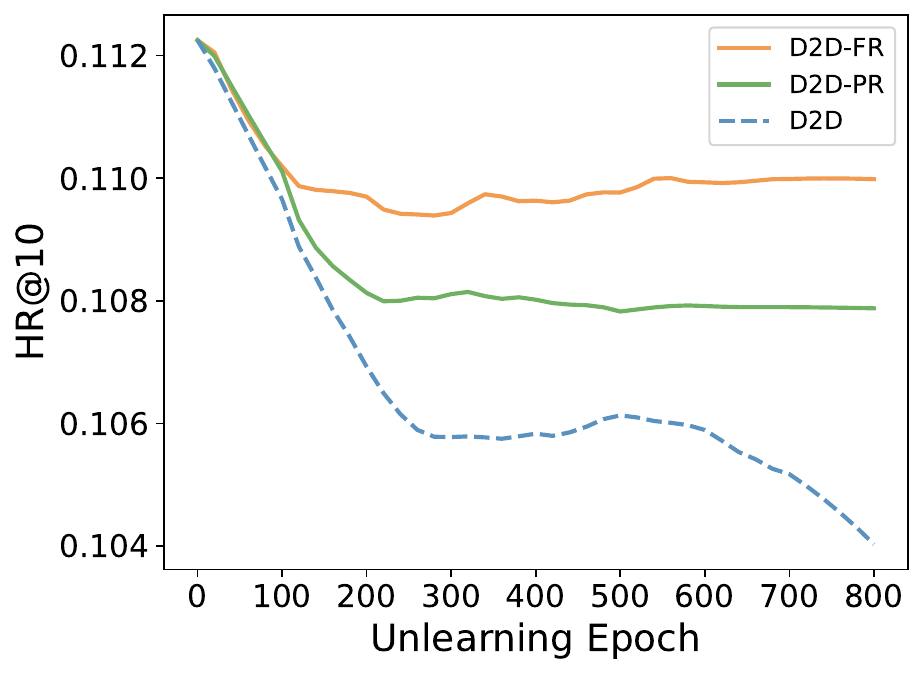} \label{fig:hr_epoch}}
\subfigure[\textbf{NDCG@10 on ML1M}]{\includegraphics[width=4.3cm]{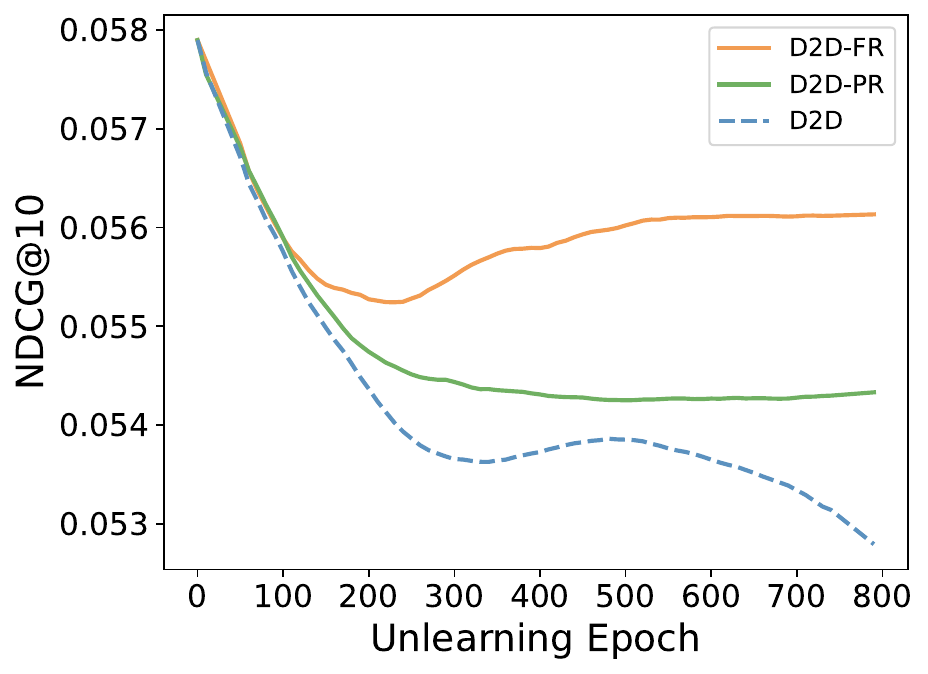} \label{fig:ndcg_epoch}}
\caption{Change in recommendation performance during unlearning, where the x-axis and y-axis represent unlearning epochs and values of metric, respectively. 
The BAcc of attackers for D2D-FR, D2D-PR, and D2D (after running 800 epochs) are 0.3334, 0.3333, and 0.3333.}
\label{fig:epoch_rec}
\end{figure}

\begin{figure}[t]
\centering
\subfigure[\textbf{$\ell_2$ loss (optimization)}]{\includegraphics[width=4.25cm]{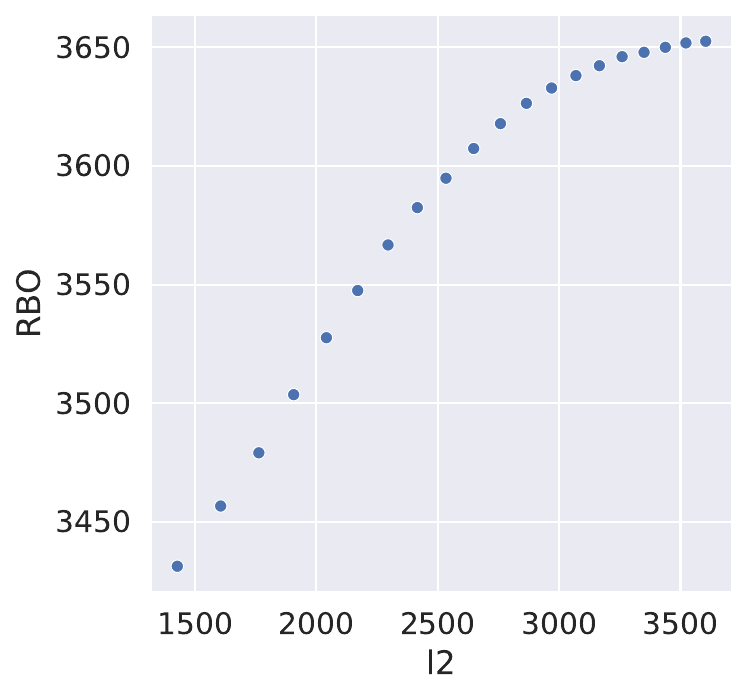} \label{fig:new_l2_cor_u}}
\subfigure[\textbf{$\ell_r$ loss (optimization)}]{\includegraphics[width=4.25cm]{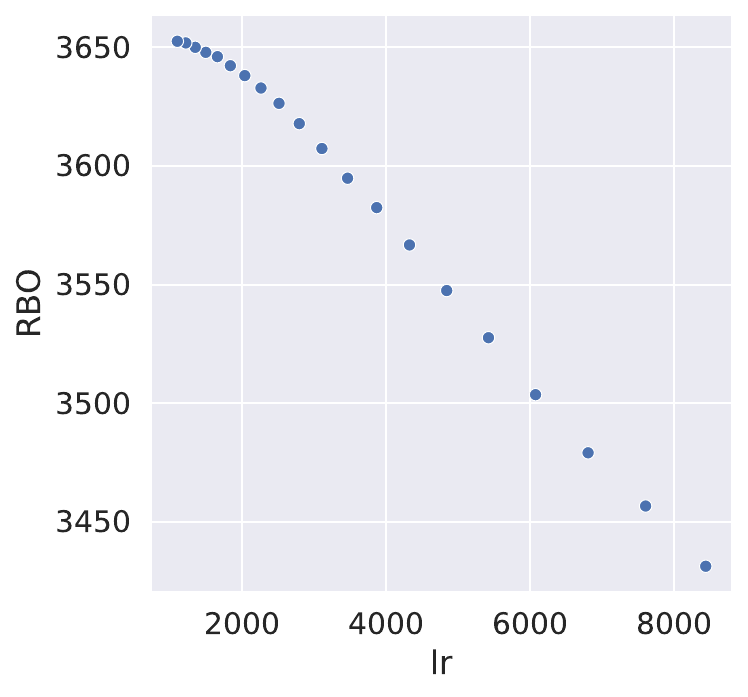} \label{fig:new_kd_cor_u}}
\caption{Correlation between two types of regularization losses and RBO (similarity in recommendation performance, specified in Section~\ref{sec:compare}) during optimization with D2D-FR, where the x-axis represents $\ell_2$ and $\ell_r$ respectively and y-axis represents RBO, each point represents a certain epoch. (a) $\ell_2$ and RBO; (b) $\ell_r$ and RBO. 
There is a notable negative correlation between RBO and $\ell_r$, but not between RBO and $l_2$.
A negative correlation indicates a valid loss measurement, as smaller loss values correspond to greater similarity in recommendation performance.}
\label{fig:corr}
\end{figure}

\begin{figure*}[t]
\centering
\subfigure[\textbf{ML-100K}]{\includegraphics[width=3.4cm]{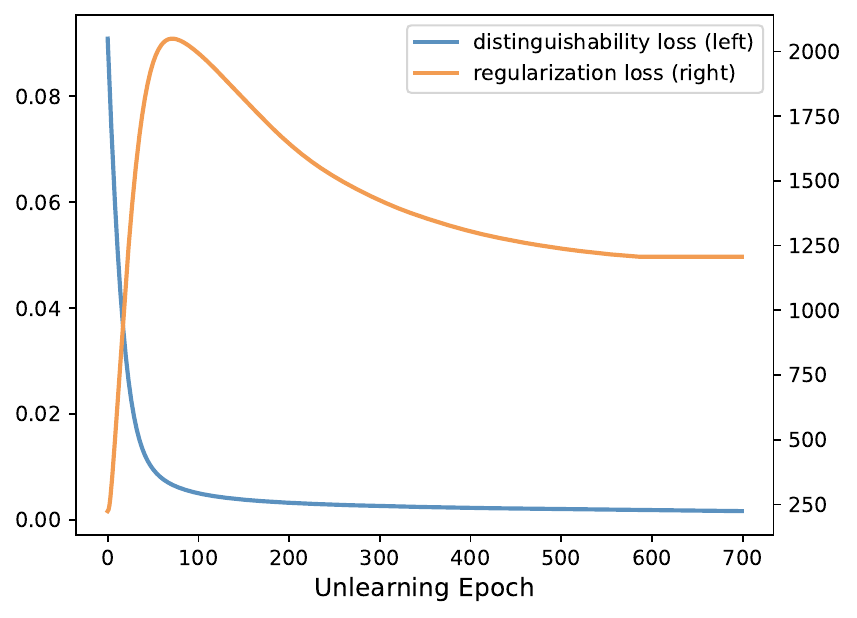}}
\subfigure[\textbf{ML-1M}]{\includegraphics[width=3.4cm]{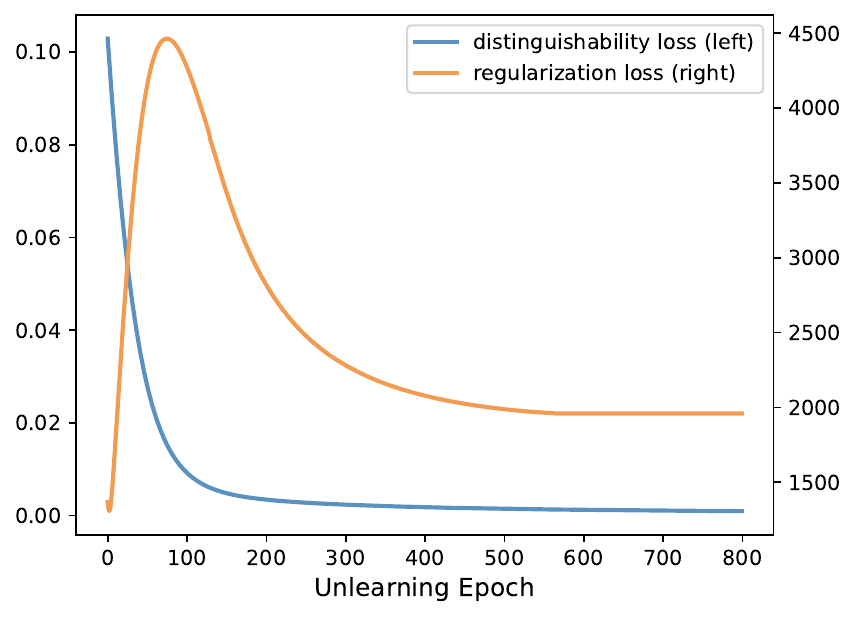}}
\subfigure[\textbf{LFM-2B}]{\includegraphics[width=3.4cm]{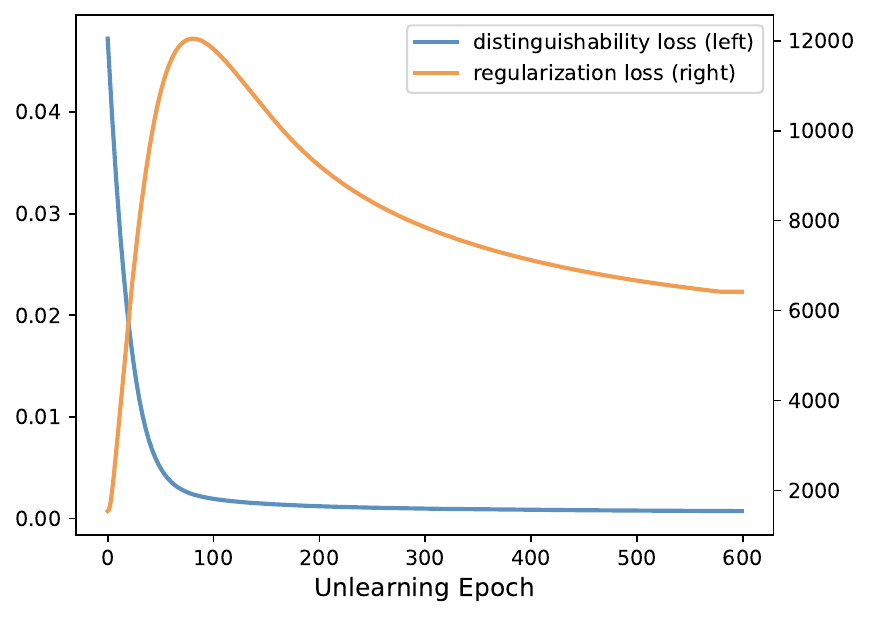}}
\subfigure[\textbf{KuaiSAR}]{\includegraphics[width=3.4cm]{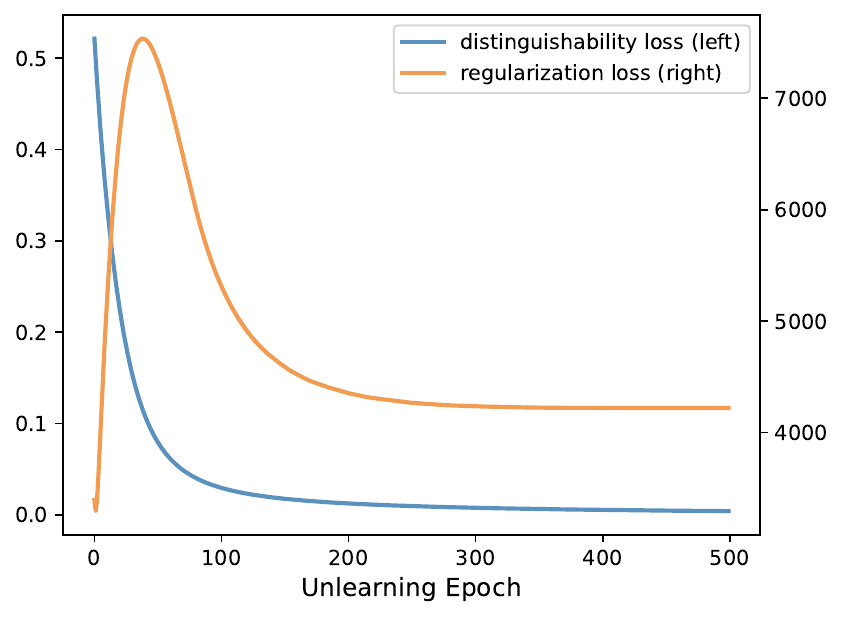}}
\caption{Change of each component's value (distinguishability loss and regularization loss) in the loss function during unlearning.}
\label{fig:loss_contra}
\end{figure*}

To understand the contribution of our proposed function-space regularization loss $\ell_r$, we compare the difference between $\ell_r$ and $\ell_2$ on preserving recommendation performance by conducting unlearning using NMF on ML-1M dataset with age as the target attribute. we report the change of recommendation performance and loss during optimization in Fig.~\ref{fig:epoch_rec} and Fig.~\ref{fig:corr} respectively. Furthermore, we analyze the potential conflict of $\ell_r$ and $\ell_u$ in Fig.~\ref{fig:loss_contra}. From these, we have the following observations:
\begin{itemize}[leftmargin=*] \setlength{\itemsep}{-\itemsep}
    \item As shown in Fig.~\ref{fig:epoch_rec}, the recommendation performance dropped significantly during the unlearning process with D2D loss, i.e., $\ell_u$. This phenomenon illustrates the necessity of introducing regularization loss to achieve \textbf{Goal \#2}. Meanwhile, compared to D2D-PR, the proposed D2D-FR is more effective to preserve the recommendation performance during optimization. 
    \item From Fig.~\ref{fig:corr}, we observe that the parameter-space regularization loss $\ell_2$ is not always negatively correlated to RBO during unlearning. In contrast, the function-space regularization loss $\ell_r$ exhibits a relatively higher correlation with RBO. Based on these, D2D-FR can search for optimal model parameters for recommendation performance after $\ell_u$ is converged. 
    \item From, Fig.~\ref{fig:loss_contra}, we observe that i) at the beginning of optimization, our proposed $\ell_r$ (regularization loss) conflicts with the D2D loss (distinguishability loss), as they move in opposite directions; ii) the D2D loss converges quickly afterward; and iii) finally, $\ell_r$ is able to align with the direction of the D2D loss, achieving a suitable balance.
\end{itemize}

\nosection{Summary} With the analysis of the unlearning process with D2D-FR, we find that our proposed D2D-FR outperforms D2D and D2D-PR in maintaining the recommendation performance, which is mainly attributed to the high correlation between $\ell_r$ and the model function during the unlearning process. 
Further analysis also finds that our proposed $\ell_r$ does not significantly conflict with the D2D unlearning loss, thereby achieving both goals concurrently.

\subsubsection{Unlearning Performance under different types of attacker (RQ6) \label{exp:rq6}}

In real-life scenarios, numerous models are available for conducting attribute inference attacks, rendering the attacker often unknown to the defenders.
To better understand the robustness of our method, we also investigate other types of attackers.
Specifically, we use gender and age as targets attribute and conduct unlearning on the ML-1M dataset.

\paragraph{Non-DNN-based Attackers.} We investigate several frequently used machine learning models in the classification task as attackers, including Decision Tree (DT), Support Vector Machine (SVM), Naive Bayes (NB), and $k$-Nearest Neighbors (KNN).
Based on the F1 score and BAcc of each attacker shown in Table~\ref{tab:multi_attack}, we have these observations:
\begin{itemize}[leftmargin=*] \setlength{\itemsep}{-\itemsep}
    \item It is obvious that our proposed D2D-PR and D2D-FR outperform Adv-InT and achieve the same unlearning performance as retrain in most scenarios, which implies that our methods can more effectively erase attribute information from the recommendation model and protect the privacy of users when confronted with unknown attacker models. Specifically, Retrain, Adv-InT, D2D-PR and D2D-FR decrease the BAcc by 34.91\%, 28.04\%, 34.26\% and 35.36\% respectively. In most cases, the BAcc after unlearning is similar to that of a random attacker.
    \item As trained to defend a specific DNN-based inference model, Adv-InT deteriorates the unlearning performance when the attacker employs non-DNN-based models. Specifically, Adv-InT decreases BAcc by 32.89\% when the attacker is MLP, whereas it decreases BAcc by 26.83\% in average when the attacker is not MLP. 
    \item The DNN-based attacker (i.e., MLP) outperforms other attackers in most scenarios due to its superiority in learning the non-linear correlation between user embeddings and the labels of target attributes.
\end{itemize}

\paragraph{Ensemble Learning-based Attackers.} Ensemble learning is a widely-used technique to improve the performance of classification models. We investigate some acknowledged ensemble learning-based attackers, including
Random Forest (RF), AdaBoost, XGBoost, and GBDT, and report the results in  Table~\ref{tab:multi_attack_ensemble}.
From it, we observe that i) our proposed D2D-FR and D2D-PR consistently outperform the compared method in terms of unlearning performance across different ensemble learning-based attackers; and ii) Comparing various attackers' performance on Original as a reference, MLP attacker still outperforms ensemble learning-based attackers.

\paragraph{MLP Attackers.} In previous experiments, we used a two-layer MLP attacker. In this experiment, we explore the impact of different MLP structures. Specifically, we investigate the number of layers in 1, 2, 3, and 4.
From Table~\ref{tab:multi_attack_mlp_layer}, we observe that the two-layer MLP achieves the best performance among all compared attackers. Additionally, increasing the number of layers cannot enhance attacking performance. We also notice that our proposed D2D-FR and D2D-PR outperform other compared methods in most cases.

\begin{table*}
\caption{Results of unlearning performance (performance of attribute inference attack) w.r.t. different types of attacker. The top results are highlighted in \textbf{bold}. InT-AU methods are represented in \texttt{typewriter font}.
}
\label{tab:multi_attack}
\centering
\resizebox{1\linewidth}{!}{
\begin{tabular}{cc|cc|cc|cc|cc|cc}
\toprule
\multirow{2}{*}{Attribute} & \multirow{2}{*}{Method} & \multicolumn{2}{c}{DT} & \multicolumn{2}{c}{KNN} & \multicolumn{2}{c}{SVM} & \multicolumn{2}{c}{NB} &  \multicolumn{2}{c}{MLP} \\
\cmidrule{3-12}
& & F1 & BAcc &  F1 & BAcc & F1 & BAcc & F1 & BAcc & F1 & BAcc\\
\midrule
\multirow{5}{*}{Gender} & Original  & 0.6255 & 0.6270 & 0.7408 & 0.7305 & 0.7585 & 0.7580 & 0.7340 & 0.7326 & 0.7602 & 0.7597 \\
& \texttt{Retrain}                  & \textbf{0.5035} & 0.5056 & 0.4895 & 0.5037 & 0.4978 & 0.4917 & 0.5153 & 0.4895 & 0.5003 & \textbf{0.5009} \\
& \texttt{Adv-InT}                  & 0.5314 & 0.5437 & 0.5663 & 0.5582 & 0.5642 & 0.5573 & 0.5734 & 0.5605 & 0.5774 & 0.5551\\
& D2D-PR                            & 0.5043 & \textbf{0.5036} & 0.5180 & 0.5121 & 0.5023 & 0.4942 & 0.5337 & 0.5105 & 0.4979 & 0.5118\\
& D2D-FR                            & 0.5067 & 0.5061 & \textbf{0.4594} & \textbf{0.4956} & \textbf{0.4748} & \textbf{0.4640} & \textbf{0.5086} & \textbf{0.4810} & \textbf{0.4944} & 0.5035 \\
\cmidrule{0-11}
\multirow{5}{*}{Age}  & Original  
          & 0.5539 & 0.4661 & 0.6563 & 0.5055  & 0.7182 & 0.6084 & 0.6614 & 0.5600 & 0.7166 & 0.6061 \\
& \texttt{Retrain} & 0.4151 & 0.3354 & 0.5025 & \textbf{0.3153} & 0.5665 & 0.3333 & 0.5664 & \textbf{0.3334} & 0.5667 & 0.3338 \\
& \texttt{Adv-InT} & 0.4355 & 0.3574 & 0.5521 & 0.3475  & 0.6036 & 0.3834 & 0.5863 & 0.3572 & 0.6125 & 0.3707 \\
& D2D-PR  & 0.4153 & \textbf{0.3350} & 0.5055 & 0.3195  & 0.5664 & 0.3333 & 0.5667 & 0.3350 & \textbf{0.5664} & 0.3334 \\
& D2D-FR  & \textbf{0.4149} & 0.3383 & \textbf{0.4975} & 0.3167 & \textbf{0.5664} & \textbf{0.3333} & \textbf{0.5662} & 0.3341 & 0.5665 & \textbf{0.3334} \\
\bottomrule
\end{tabular}
}
\end{table*}

\begin{table*}
\caption{Results of unlearning performance (performance of attribute inference attack) w.r.t. different types of ensemble learning-based attacker. The top results are highlighted in \textbf{bold}. InT-AU methods are represented in \texttt{typewriter font}.
}
\label{tab:multi_attack_ensemble}
\centering
\resizebox{0.833\linewidth}{!}{
\begin{tabular}{cc|cc|cc|cc|cc}
\toprule
\multirow{2}{*}{Attribute} & \multirow{2}{*}{Method} & \multicolumn{2}{c}{RF} & \multicolumn{2}{c}{AdaBoost} & \multicolumn{2}{c}{XGBoost} & \multicolumn{2}{c}{GBDT} \\
\cmidrule{3-10}
& & F1 & BAcc &  F1 & BAcc & F1 & BAcc & F1 & BAcc \\
\midrule
\multirow{5}{*}{Gender} & Original  & 0.7313 & 0.7325 & 0.7143 & 0.7153 & 0.7392 & 0.7382 & 0.7452 & 0.7426 \\
& \texttt{Retrain}      & 0.4931 & 0.5097 & 0.5023 & 0.5031 & 0.4913 & 0.4975 & 0.5134 & 0.5107 \\
& \texttt{Adv-InT}      & 0.5336 & 0.5579 & 0.5325 & 0.5602 & 0.5367 & 0.5528 & 0.5453 & 0.5514 \\
& D2D-PR                & 0.4827 & 0.5066 & 0.4961 & 0.4973 & \textbf{0.4884} & \textbf{0.4874} & 0.5132 & 0.5089 \\
& D2D-FR                & \textbf{0.4793} & \textbf{0.5017} & \textbf{0.4918} & \textbf{0.4971} & 0.5052 & 0.5149 & \textbf{0.5123} & \textbf{0.5027}  \\
\cmidrule{0-9}
\multirow{5}{*}{Age}  & Original  
                   & 0.6797& 0.5238& 0.6841& 0.5751& 0.7013& 0.5868& 0.6992& 0.5733\\
& \texttt{Retrain} & 0.5701& 0.3365& 0.5575& 0.3343& 0.5266& 0.3313& 0.5585& 0.3372\\
& \texttt{Adv-InT} & 0.5831& 0.3552& 0.5762& 0.3545& 0.5479& 0.3501& 0.5743& 0.3617\\
& D2D-PR           & \textbf{0.5645}& 0.3347& 0.5543& 0.3331& \textbf{0.5230}& \textbf{0.3279}& 0.5599& 0.3350 \\
& D2D-FR           & 0.5673& \textbf{0.3342}& \textbf{0.5538}& \textbf{0.3314}& 0.5241& 0.3291& \textbf{0.5534}& \textbf{0.3305}\\
\bottomrule
\end{tabular}
}
\end{table*}

\begin{table*}
\caption{Results of unlearning performance (performance of attribute inference attack) w.r.t. different types of MLP-based attacker. The top results are highlighted in \textbf{bold}. InT-AU methods are represented in \texttt{typewriter font}. the dimensions of layers are $\{d_{out}\}$, $\{100, d_{out}\}$, $\{100, 64, d_{out}\}$, $\{100,64,32,d_{out}\}$, where $d_{out}$ denotes the count of attribute categories.
}
\label{tab:multi_attack_mlp_layer}
\centering
\resizebox{0.833\linewidth}{!}{
\begin{tabular}{cc|cc|cc|cc|cc}
\toprule
\multirow{2}{*}{Attribute} & \multirow{2}{*}{Method} & \multicolumn{2}{c}{Layer=1} & \multicolumn{2}{c}{Layer=2} & \multicolumn{2}{c}{Layer=3} & \multicolumn{2}{c}{Layer=4} \\
\cmidrule{3-10}
& & F1 & BAcc &  F1 & BAcc & F1 & BAcc & F1 & BAcc \\
\midrule
\multirow{5}{*}{Gender} & Original  & 0.7522 & 0.7539 & 0.7608 & 0.7607 & 0.7373 & 0.7363 & 0.7167 & 0.7178 \\
& \texttt{Retrain}      & 0.4835& 0.4973& 0.4902& 0.5061& 0.5077& 0.5052& 0.4921& 0.5009\\
& \texttt{Adv-InT}      & 0.5251& 0.5377& 0.5279& 0.5343& 0.5319& 0.5383& 0.5226& 0.5349\\
& D2D-PR                & \textbf{0.4816}& 0.4959 & 0.4893 & 0.5063 & 0.5097& 0.5065& 0.4848& 0.4972\\
& D2D-FR                & 0.4863& \textbf{0.4642}& \textbf{0.4817}& \textbf{0.4907}& \textbf{0.5073}& \textbf{0.5004}& \textbf{0.4835}& \textbf{0.4892}\\
\cmidrule{0-9}
\multirow{5}{*}{Age}  & Original  
                   & 0.7124& 0.5890& 0.7183& 0.6075& 0.7157& 0.6177& 0.7126& 0.6076\\
& \texttt{Retrain} & 0.5669& 0.3335& 0.5668& 0.3343& 0.5683& 0.3355& 0.5672& 0.3347\\
& \texttt{Adv-InT} & 0.6108& 0.3775& 0.6059& 0.3747& 0.5989& 0.3761& 0.5973& 0.3776\\
& D2D-PR           & 0.5666& 0.3336& 0.5666& \textbf{0.3336}& 0.5664& 0.3333& 0.5664& 0.3333\\
& D2D-FR           & \textbf{0.5666}& \textbf{0.3335}& \textbf{0.5565}& 0.3337& \textbf{0.5664}& \textbf{0.3333}& \textbf{0.5664}& \textbf{0.3333}\\
\bottomrule
\end{tabular}
}
\end{table*}

\section{Conclusions and Future Work}\label{sec:con}

In this paper, following our previous work~\cite{li2023making}, we study the Post-Training Attribute Unlearning (PoT-AU) problem in recommender systems, which aims to protect users' attribute information instead of input data.
%
%
There are two goals in the PoT-AU problem, i.e., making attributes indistinguishable, and maintaining comparable recommendation performance.
To achieve the above two goals, we propose a two-component loss function, which consists of distinguishability loss and regularization loss, to optimize model parameters.
Our previous work focuses on binary-class attributes. In this paper, we expand the applicability to the multi-class scenario. 
\lyy{To the best of our knowledge, this is the first work to explore the multi-class scenario in attribute unlearning, thereby enhancing the overall completeness and real-world applicability of our previous research.}
Specifically, we further improve the efficiency of distributional distinguishability loss in the multi-class scenario, and introduce a function-space regularization loss to directly preserve recommendation performance.
We conduct extensive experiments on four real-world datasets to evaluate the effectiveness of our proposed methods.
The results demonstrate that our newly proposed D2D-FR outperforms all compared methods, including our previous work (i.e., D2D-PR).
%
%

In this work, we focus on the system-wise attribute unlearning, i.e., conducting unlearning for all users in the system.
In future research, we plan to investigate user-wise attribute unlearning. In this scenario, only the parameters of users who request attribute unlearning will be updated, while maintaining comparable overall recommendation performance.

\begin{acks}
This work was supported in part by the ``Ten Thousand Talents Program'' of Zhejiang Province for Leading Experts (No. 2021R52001), and the National Natural Science Foundation of China (No. 72192823).
\end{acks}

\bibliographystyle{ACM-Reference-Format}





\end{document}